\UseRawInputEncoding
\documentclass[preprint,aps,12pt,showpacs,nofootinbib,tightenlines]{revtex4}
\usepackage{mathrsfs}
\usepackage{amsmath}
\usepackage{amssymb}
\usepackage{epsfig}
\usepackage{epstopdf}
\usepackage{graphicx}
\usepackage{subfigure}
\usepackage{booktabs}
\usepackage{float}
\usepackage{amsmath}
\usepackage{color}
\usepackage{multirow}
\def\be{\begin{eqnarray}}
\def\en{\end{eqnarray}}
\def\non{\nonumber\\}

\begin{document}
\title{
 Semileptonic $B_{(s)}$ meson decays to $D_{0}^{\ast}(2300) ,D_{s0}^{\ast}(2317) , D_{s1}(2460), D_{s1}(2536), D_{1}(2420)$ and $D_{1}(2430)$ within the covariant light-front approach  }
\author{ You-Ya Yang,$^1$ Zhi-Qing Zhang,$^{1,}$\footnote{ zhangzhiqing@haut.edu.cn (corresponding author)} and Hao Yang$^1$}
\affiliation{ \it \small $^1$  School of Physics, Henan University of Technology,
 Zhengzhou, Henan 450001, China}
\date{\today}
\begin{abstract}
In this work, we investigate the semileptonic decays of $B_{(s)}$ meson to $D_{0}^{\ast}(2300)$, $D_{s0}^{\ast}(2317)$, $D_{s1}(2460)$, $D_{s1}(2536)$, $D_{1}(2420)$ and $D_{1}(2430)$ in the covariant light-front quark model (CLFQM). We combine the helicity amplitudes via the corresponding form factors to obtain the branching ratios of the semileptonic decays $B_{(s)} \to  D^{**}_{(s)}\ell \nu_\ell$ with $D^{**}_{(s)}$ referring to a P-wave exicted charmed meson $D_{0}^{\ast}(2300)$, $D_{s0}^{\ast}(2317)$, $D_{s1}(2460)$, $D_{s1}(2536)$, $D_{1}(2420)$ or $D_{1}(2430)$ and  $\ell=e, \mu, \tau$.  Furthermore, we also take into account another two physical observables, namely the longitudinal polarization fraction $f_L$ and the forward-backward asymmetry $A_{FB}$. Most of our predictions are comparable to the results given by other theoretical approaches and the present available data. The branching ratios of the semileptonic decay channels $B_{s} \to D_{s1}(2460) \ell \nu_\ell$ and $B \to D_{1}(2420) \ell \nu_\ell$ are larger than those of the semileptonic decays $B_{s} \to D_{s1}(2536) \ell \nu_\ell$ and $B \to D_{1}(2430) \ell \nu_\ell$, respectively. We find that the long-standing '$1/2$ vs $3/2$ puzzle' in the decays $B^{+}\to \bar{D}_1^{(\prime)0} \ell^{\prime+}\nu_{\ell^\prime}$ $(\ell^\prime=e,\mu)$ can be solved by taking some negative mixing angle $\theta_s$ values within a range from $-30.3^\circ$ to $-24.9^\circ$, corresponding to $\theta$ of about $5^\circ\sim10.4^\circ$. While Belle collaboration updated their measurements for the decays $B^{0}\to D^{*-}_0 \ell^{\prime+}\nu_{\ell^\prime}$ with only a small upper limit $Br(B^{0}\to D^{*-}_0 \ell^{\prime+}\nu_{\ell^\prime})<0.44\times10^{-3}$ obtained, which is much larger than most theoretical predictions and causes a new puzzle.
\end{abstract}

 \vspace{1cm}

\maketitle

\section{Introduction}\label{intro}
In the conventional quark model, these P-wave orbitally excited charmed mesons $D_{0}^{\ast}(2300)$, $D_{s0}^{\ast}(2317)$, $D_{s1}(2460)$, $D_{s1}(2536)$, $D_{1}(2420)$ and $D_{1}(2430)$ can be view as constituent quark-antiquark pairs. They are usually classified according to the quantum numbers $^{(2S+1)}L_J$: the scalar mesons $D_{0}^{\ast}(2300)$ and $D_{s0}^{\ast}(2317)$ correspond to $^3P_0$. While there exist two different kinds of axial-vector mesons, namely $^1P_1$ and $^3P_1$, which can undergo mixing when the two constituent quarks are different. In the heavy quark limit, the heavy quark spin and the total angular momentum of the light quark are good quantum numbers, it is more convenient to use the $L^j_J$ configurations \footnote{$j(L)$ being the total (orbital) angular momentum of the light quark.} to classify them: the scalar mesons $D_{0}^{\ast}(2300)$ and $D_{s0}^{\ast}(2317)$ belong to $P^{1/2}_0$, $D_1(2430)(D_{s1}(2536))$ and  $D_1(2420)(D_{s1}(2460))$ correspond to $P^{3/2}_1$ and $P^{1/2}_1$, respectively. However, beyond the heavy quark limit, there is a mixing between $P^{3/2}_1$ and $P^{1/2}_1$, denoted by $D^{3/2}_{1}$ and $D^{1/2}_{1}$, respectively\footnote{Here we take the physical states $D_1(2430)$ and  $D_1(2420)$ as an example to explain, it is similar to the states $D_{s1}(2536)$ and  $D_{s1}(2460)$.}, that is
\begin{eqnarray}
	|D_{1}(2420)\rangle &=&|D_{1}^{1 / 2}\rangle  \sin \theta_s+|D_{1}^{3 / 2}\rangle  \cos \theta_s, \nonumber\\
	|D_{1}(2430)\rangle &=&-|D_{1}^{3 / 2}\rangle  \sin \theta_s+|D_{1}^{1 / 2}\rangle  \cos \theta_s.
	\label{mixing1}
\end{eqnarray} 
While the states $D_{1}^{1/2}$ and $D_{1}^{3/2}$ are expected to be a mixture of states $^1P_{1}$ and $^3P_{1}$ denoted by $^1D_{1}$ and $^3D_{1}$, respectively,
\begin{eqnarray}
	|D^{3/2}_{1}\rangle &=& \sqrt{\frac{2}{3}} |^1D_{1}\rangle +\sqrt{\frac{1}{3}}
	|^3D_{1}\rangle,\nonumber\\
	|D^{1/2}_{1}\rangle &=& -\sqrt{\frac{1}{3}} |^1D_{1}\rangle +\sqrt{\frac{2}{3}}
	|^3D_{1}\rangle.\label{mixing2}
\end{eqnarray}
Combining Eq. (\ref{mixing1}) and Eq. (\ref{mixing2}), one can find that the physical states $D_1(2420)$ and $D_1(2430)$ can be written as 
\be
|D_{1}(2420)\rangle &=& |^1D_{1}\rangle \cos \theta+|^3D_{1}\rangle\sin \theta , \nonumber\\
|D_{1}(2430)\rangle &=& -|^1D_{1}\rangle \sin \theta+|^3D_{1}\rangle\cos \theta.
\label{mixing3}
\en
where $\theta_s=7^\circ$ and $\theta=\theta_s+35.3^\circ$ \cite{zhw}.
There exist many puzzles in these several 
P-wave excited states, such as the low mass puzzle for the states $D^*_0(2300), D^*_{s0}(2317)$ and $D_{s1}(2460)$ \cite{belle,quark,babar2,godfrey2}, the SU(3) mass hierarchy puzzle between $D^*_0(2300)$ and $D^*_{s0}(2317)$, large width difference between them \cite{Gubernari}, especially, the long-standing '$1/2$ vs $3/2$ puzzle' \cite{morenas,bigi,scora,colangelo}, that is the theoretical predictions for the branching ratios of semileptonic B decays into $D^{1/2+}$ are much smaller than those into $D^{3/2+}$, which conflicts with the experimental measurements \cite{belle,babar2,Belle:2022yzd}. 
These unexpected  disparities between theory and experiment have sparked many explanations about their inner structures, such as the molecular states, the compact tetraquark states, the states of $\bar cs$ mixed with four-quark states, and so on \cite{guo,cleven,close,guofk,lutz,lutz2,ylma,maiani,wangzg,hycheng2,yqchen,kim,bardeen,nowak,browder,vijande}.

In this paper we investigate the semileptonic $B_{(s)}$ meson decays to $D_{0}^{\ast}(2300)$, $D_{s0}^{\ast}(2317)$, $D_{s1}(2460)$, $D_{s1}(2536)$, $D_{1}(2420)$ and $D_{1}(2430)$ by using the covariant light-front quark model (CLFQM). 
For the semileptonic decays, the hadronic transition matrix element between the initial
and final mesons is most crucial for theoretical calculations, which can be characterized
by several form factors. The form factors can be extracted from data or relied on some non-perturbative methods. The $B_{(s)}\to D^{**}_{(s)}$ transition form factors were initially calculated in the improved version of the Isgur-Scora-Ginstein-Wise (ISGW) quark model, the so-called ISGW2 \cite{hycheng}. Some of them were calculated using the CLFQM \cite{Cheng}, QCD sum rules (QCDSR) \cite{Y.B.,M.Q.,Zuo:2023ksq} and  light-cone sum rules (LCSRs) \cite{Gubernari}. Additionally, with the available experimental data as inputs, the form factors of the $B_{(s)}$ to these excited charmed meson transitions and the corresponding semileptonic decays were also investigated based on the heavy quark effective theory (HQET), including the next-to-leading order corrections of heavy quark expansion and new physics (NP) effects \cite{A.K.1,A.K.2,F.U.,F.U.1}. 
As one of the popular non-perturbative methods, the CLFQM has been used successfully to study the form factors \cite{Cheng,Cheng1,Hwang,Lu,Wang}.  Based on the form factors and helicity formalisms, we
also calculate another two physical observables: the forward-backward asymmetry $A_{FB}$ and the longitudinal polarization fraction $f_L$, respectively.

The arrangement of this paper is as follows: In Section II, the formalism of the CLFQM, the hadronic matrix
elements and the helicity amplitudes combined via form factors are presented. The numerical results for the $B_{(s)}$ meson to $D_{0}^{\ast}(2300)$, $D_{s0}^{\ast}(2317)$, $D_{s1}(2460)$, $D_{s1}(2536)$, $D_{1}(2420)$ and $D_{1}(2430)$
transition form factors, the branching ratios, the forward-backward asymmetries $A_{FB}$ and
the longitudinal polarization fractions $f_L$ for the corresponding decays are presented in Section III.  In addition, the detailed numerical analysis and
discussion, including comparisons with the data and other model calculations, are carried out. The conclusions are presented in the final part.

\section{Formalism}\label{form}
\subsection{The covariant light-front quark model}
Under the covariant light-front quark model, the light-front coordinates of a momentum $p$ are defined as $p=(p^-,p^+,p_\perp)$ with
$p^\pm=p^0\pm p_z$ and $p^2=p^+p^--p^2_\perp$. If the momenta of the quark and antiquark
with masses $m_{1}^{\prime(\prime\prime)}$ and $m_2$ in the incoming (outgoing) meson are denoted as $p_{1}^{\prime(\prime\prime)}$
and $p_{2}$, respectively, the momentum of the incoming (outgoing) meson with mass $M^\prime(M^{\prime\prime})$ can be written as $P^\prime=p_1^\prime+p_2 (P^{\prime\prime}=p_1^{\prime\prime}+p_2)$.
Here, we use the same notation
as those in Refs. \cite{Jaus,Cheng} and $M^\prime$ refers to $m_{B}$ for $B$ meson decays.
These momenta can be related each other through the internal variables $(x_{i},p{\prime}_{\perp})$
\be
p_{1,2}^{\prime+}=x_{1,2} P^{\prime+}, \quad p_{1,2 \perp}^{\prime}=x_{1,2} P_{\perp}^{\prime} \pm p_{\perp}^{\prime},
\en
with $x_{1}+x_{2}=1$. Using these internal variables,
we can define some quantities for the incoming meson which will be used in the following calculations
\be
M_{0}^{\prime 2} &=&\left(e_{1}^{\prime}+e_{2}\right)^{2}=\frac{p_{\perp}^{\prime 2}+m_{1}^{\prime 2}}{x_{1}}
+\frac{p_{\perp}^{2}+m_{2}^{2}}{x_{2}}, \quad \widetilde{M}_{0}^{\prime}=\sqrt{M_{0}^{\prime 2}-\left(m_{1}^{\prime}-m_{2}\right)^{2}}, \non
e_{i}^{(\prime)} &=&\sqrt{m_{i}^{(\prime) 2}+p_{\perp}^{\prime 2}+p_{z}^{\prime 2}}, \quad \quad p_{z}^{\prime}
=\frac{x_{2} M_{0}^{\prime}}{2}-\frac{m_{2}^{2}+p_{\perp}^{\prime 2}}{2 x_{2} M_{0}^{\prime}},\en
where the kinetic invariant mass of the incoming meson $M^\prime_0$ can be expressed as the energies of the quark and the antiquark
$e^{(\prime)}_i$. It is similar to the case of the outgoing meson.
\begin{figure}[H]
\centering \subfigure{
\begin{minipage}{5cm}
\centering
\includegraphics[width=5cm]{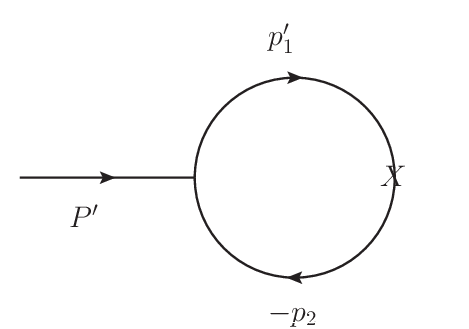}
\end{minipage}}
\subfigure{
\begin{minipage}{6cm}
\centering
\includegraphics[width=6cm]{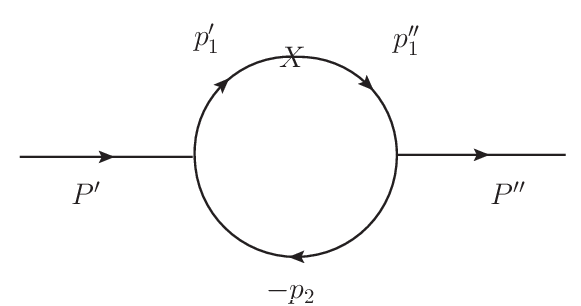}
\end{minipage}}
\caption{Feynman diagrams for $B_{(s)}$ decay (left) and transition
(right) amplitudes, where $P^{\prime(\prime\prime)}$ is the
incoming (outgoing) meson momentum, $p^{\prime(\prime\prime)}_1$
is the quark momentum and $p_2$ is the anti-quark momentum. The X in the diagrams denotes the vector or axial-vector transition vertex.}
\label{feyn}
\end{figure}

To calculate the amplitudes for the transition form factors,
we need the Feynman rules for the meson-quark-antiquark vertices $i\Gamma^\prime_M$ \footnote{In the following we take the transitions $B\to D^*_{0}$ and $B\to ^3D_{1}, ^1D_{1}$ as examples. It is similar for the transitions $B_s\to D^*_{s0}$ and $B_s\to ^3D_{s1}, ^1D_{s1}$. From now on, we will use $D^*_0, D^*_{s0}, D_{s1}, D_{s1}^{\prime}, D_{1}$ and $D_{1}^{\prime}$ to represent $D^*_0(2300), D^*_{s0}(2317), D_{s1}(2460), D_{s1}(2536), D_{1}(2420)$ and $D_{1}(2430)$, respectively, for simplicity. }, which are listed as
\be
i \Gamma_{D^*_0}^{\prime}&=&-i H_{D^*_0}^{\prime},\\
i \Gamma_{\;^3D_{1}}^{\prime}&=&-i H_{\;^3D_{1}}^{\prime}\left[\gamma_{\mu}+\frac{1}{W_{\;^3D_{1}}^{\prime}}\left(p_{1}^{\prime}-p_{2}\right)_{\mu}\right] \gamma_{5},\\
i \Gamma_{\;^1D_{1}}^{\prime}&=&-i H_{\;^1D_{1}}^{\prime}\left[\frac{1}{W_{\;^1D_{1}}^{\prime}}\left(p_{1}^{\prime}-p_{2}\right)_{\mu}\right] \gamma_{5}.
\en
The form factors of the transitions
$B\to D^*_0$  and $B\to \ ^iD_{1} (i=1, 3)$ induced by the vector and aixal-vector currents
are defined as
\be
\left\langle D^*_0 \left(P^{\prime
\prime}\right)\left|A_{\mu}\right|
B\left(P^{\prime}\right)\right\rangle&=&i\left[u_+(q^2)P_\mu+u_-(q^2)q_\mu\right],  \label{form11} \\
\left\langle \; ^iD_{1} \left(P^{\prime
\prime},\varepsilon\right)\left|A_{\mu}\right|
B\left(P^{\prime}\right)\right\rangle&=-&q(q^2)\epsilon_{\mu\nu\alpha\beta}\varepsilon^{*\nu}P^\alpha q^\beta,  \label{form22}\\
\left\langle \; ^iD_{1}\left(P^{\prime
\prime},\varepsilon\right)\left|V_{\mu}\right|
B\left(P^{\prime}\right)\right\rangle&=&i\left\{l(q^2)\varepsilon_\mu^*+\varepsilon^*\cdot P\left[P_\mu c_+(q^2)+q_\mu c_-(q^2)\right]\right\}. \label{form33}
\en
In calculations, the Bauer-Stech-Wirbel (BSW) \cite{bsw} transition form factors are more frequently used and defined by
\begin{footnotesize}
\begin{eqnarray}
\left\langle D^*_0 \left(P^{\prime
	\prime}\right)\left|A_{\mu}\right|
B\left(P^{\prime}\right)\right\rangle&=&\left(P_{\mu}-\frac{m_{B}^{2}-m_{D^*_0}^{2}}{q^{2}}
q_{\mu}\right) F_{1}^{B
	D^*_0}\left(q^{2}\right)+\frac{m_{B}^{2}-m_{D^*_0}^{2}}{q^{2}} q_{\mu}
F_{0}^{B D^*_0}\left(q^{2}\right),\;\;\;\;\;\;\;\;\\
\left\langle ^iD_{1} \left(P^{\prime \prime}, \varepsilon^{\mu *}\right)\left|V_{\mu}\right| B\left(P^{\prime}\right)\right\rangle
&=&-i\left\{\left(m_{B}-m_{^iD_{1}}\right) \varepsilon_{\mu}^{ *} V_{1}^{B \ ^iD_{1}}\left(q^{2}\right)
-\frac{\varepsilon^{ *} \cdot P}{m_{B}-m_{^iD_{1}}} P_{\mu} V_{2}^{B \ ^iD_{1}}\left(q^{2}\right)\right.\non &&
\left.-2 m_{^iD_{1}} \frac{\varepsilon^{ *} \cdot
P}{q^{2}} q_{\mu}\left[V_{3}^{B \
^iD_{1}}\left(q^{2}\right)-V_{0}^{B \
^iD_{1}}\left(q^{2}\right)\right]\right\},\label{bctoa1}\\
\left\langle ^iD_{1}\left(P^{\prime \prime}, \varepsilon^{\mu*}\right)\left|A_{\mu}\right| B\left(P^{\prime}\right)\right\rangle
&=&-\frac{1}{m_{B}-m_{^iD_{1}}} \epsilon_{\mu \nu \alpha \beta} \varepsilon^{ * \nu} P^{\alpha} q^{\beta} A^{B\ ^iD_{1}}\left(q^{2}\right),
\end{eqnarray}
\end{footnotesize}
where $P=P'+P'', q=P'-P''$ and the convention $\epsilon_{0123}=1$ is adopted.

To smear the singularity at $q^2=0$ in Eq. (\ref{bctoa1}), the following relations are required
\be
V^{B \ ^iD_{1}}_3(0)&=&V^{B\ ^iD_{1}}_0(0), \\
V^{B\ ^iD_{1}}_3(q^2)&=&\frac{m_{B}-m_{^iD_{1}}}{2m_{^iD_{1}}}V^{B\ ^iD_{1}}_1(q^2)-\frac{m_{B}+m_{^iD_{1}}}{2m_{^iD_{1}}}V^{B\ {^iD_{1}}}_2(q^2).
\en
These two kinds of form factors are related to each other via
\be
F^{BD^*_0}_1(q^2)&=&-u_+(q^2),F^{BD^*_0}_0(q^2)=-u_+(q^2)-\frac{q^2}{q\cdot P}u_-(q^2),\label{relations}\\
A^{B\ ^iD_{1}}(q^2)&=&-(m_{B}-m_{^iD_{1}})q(q^2), V^{B\ ^iD_{1}}_1(q^2)=-\frac{l(q^2)}{m_{B}-m_{^iD_{1}}},\label{relationa1}\\
V^{B\ ^iD_{1}}_2(q^2)&=&(m_{B}-m_{^iD_{1}})c_+(q^2),V^{B\ ^iD_{1}}_3(q^2)-V^{B\ ^iD_{s1}}_0(q^2)=\frac{q^2}{2m_{^iD_{1}}}c_-(q^2).\label{relationa2}
\en
For the general $B\rightarrow M$ transitions with $M$ being a scalar or axial-vector meson, the decay amplitude at the lowest order is \cite{Cheng:2003sm}
\be
\mathcal{M}^{B M}=-i^{3} \frac{N_{c}}{(2 \pi)^{4}} \int d^{4} p_{1}^{\prime} \frac{H_{B}^{\prime}\left(H_{M}^{\prime \prime}\right)}
{N_{1}^{\prime} N_{1}^{\prime \prime} N_{2}} S_{\mu}^{B M},\label{amplitude}
\en
where $N_{1}^{\prime(\prime \prime)}=p_{1}^{\prime(\prime \prime) 2}-m_{1}^{\prime (\prime\prime) 2}$ and $ N_{2}=p_{2}^{2}-m_{2}^{2} $ arise
from the quark propagators. For our considered transitions $B\rightarrow D^*_0$ and $B\to \ ^1D_{1},\ ^3D_{1}$,  
the traces $S_{\mu}^{BD^*_0}, S_{\mu \nu}^{B\;^1D_{1}}$ and $S_{\mu \nu}^{B\;^3D_{1}}$ can be directly written out by using the Lorentz contraction as follows
\begin{footnotesize}
\begin{eqnarray}
S_{\mu}^{B D^*_0} &=& Tr\left[\left(\not p_{1}^{\prime \prime}+m_{1}^{\prime \prime}\right) \gamma_{\mu} \gamma_{5}\left(\not p_{1}^{\prime}
+m_{1}^{\prime}\right) \gamma_{5}\left(-\not p_{2}+m_{2}\right)\right],\\
S_{\mu \nu}^{B\;^1D_{1}} &=&\left(S_{V}^{B \;^1D_{1}}-S_{A}^{B \;^1D_{1}}\right)_{\mu \nu} \non
&=&\operatorname{Tr}\left[\left(-\frac{1}{W_{\;^1D_{1}}^{\prime \prime}}\left(p_{1}^{\prime \prime}-p_{2}\right)_{\nu}\right) \gamma_{5}\left(\not p_{1}^{\prime \prime}
+m_{1}^{\prime \prime}\right)\left(\gamma_{\mu}-\gamma_{\mu} \gamma_{5}\right)\left(\not p_{1}^{\prime}+m_{1}^{\prime}\right) \gamma_{5}\left(-\not p_{2}+m_{2}\right)\right],\;\;\;\label{btoa1}\\
S_{\mu \nu}^{B \;^3D_{1}} &=&\left(S_{V}^{B \;^3D_{1}}-S_{A}^{B \;^3D_{1}}\right)_{\mu \nu} \non
&=&\operatorname{Tr}\left[\left(\gamma_{\nu}-\frac{1}{W_{\;^3D_{1}}^{\prime \prime}}\left(p_{1}^{\prime \prime}-p_{2}\right)_{\nu}\right) \gamma_{5}\left(\not p_{1}^{\prime \prime}
+m_{1}^{\prime \prime}\right)\left(\gamma_{\mu}-\gamma_{\mu} \gamma_{5}\right)\left(\not p_{1}^{\prime}+m_{1}^{\prime}\right) \gamma_{5}\left(-\not p_{2}+m_{2}\right)\right].\;\;\;\label{btoa3}
\end{eqnarray}
\end{footnotesize}
The form factors can be obtained by matching the coefficients listed in Eqs. (\ref{form11})-(\ref{form33}) with the corresponding amplitudes given Eq. (\ref{amplitude}). The specific expressions for these transition form factors are collected in Appendix B. It is noticed that the form factors of the transitions $B\to D_{1}$ and $B\to D_{1}^{\prime}$ can be obtained from those of the transitions $B\to \ ^1D_{1}$ and $B\to \ ^3D_{1}$ through Eq. (\ref{mixing3}).

\subsection{Wave functions and decay constants }
The light-front wave functions (LFWFs) are needed in the form factor calculations. Although the LFWFs can be derived from solving the relativistic Schr$\ddot{o}$dinger equation theoretically, it is difficult to obtain their exact solutions in many cases. Consequently, we will use the phenomenological Gaussian-type wave functions in this work,
\be
\varphi^{\prime} &=&\varphi^{\prime}\left(x_{2}, p_{\perp}^{\prime}\right)=4\left(\frac{\pi}{\beta^{\prime 2}}\right)^{\frac{3}{4}}
\sqrt{\frac{d p_{z}^{\prime}}{d x_{2}}} \exp \left(-\frac{p_{z}^{\prime 2}+p_{\perp}^{\prime 2}}{2 \beta^{\prime 2}}\right),\non
\varphi_{p}^{\prime} &=&\varphi_{p}^{\prime}\left(x_{2}, p_{\perp}^{\prime}\right)=\sqrt{\frac{2}{\beta^{\prime 2}}} \varphi^{\prime},
\quad \frac{d p_{z}^{\prime}}{d x_{2}}=\frac{e_{1}^{\prime} e_{2}}{x_{1} x_{2} M_{0}^{\prime}},\label{betap}
\en
where the parameter $\beta^\prime$ describes the momentum distribution and is approximately of order $\Lambda_{QCD}$. It can be usually determined by the decay constants through the following analytic expressions \cite{Jaus,Cheng},
\be
f_{D^*_0}&=&\frac{N_{c}}{16 \pi^{3}} \int d x_{2} d^{2} p_{\perp}^{\prime} \frac{h_{D^*_0}^{\prime}}{x_{1} x_{2}\left(M^{\prime 2}-M_{0}^{\prime 2}\right)}
4\left(m_{1}^{\prime} x_{2}-m_{2} x_{1}\right),\\
f_{\;^3D_{1}}&=&-\frac{N_{c}}{4 \pi^{3} M^{\prime}}  \int d x_{2} d^{2} p_{\perp}^{\prime} \frac{h_{^3D_{1}}^{\prime}}{x_{1} x_{2}\left(M^{\prime 2}-M_{0}^{\prime 2}\right)}\non &&
\times\left[x_{1} M_{0}^{\prime 2}-m_{1}^{\prime}\left(m_{1}^{\prime}+m_{2}\right)-p_{\perp}^{\prime 2}-\frac{m_{1}^{\prime}-m_{2}}{w_{\;^3D_{1}}^{\prime}} p_{\perp}^{\prime 2}\right],\\
f_{\;^1D_{1}}&=&\frac{N_{c}}{4 \pi^{3} M^{\prime}} \int d x_{2} d^{2} p_{\perp}^{\prime} \frac{h_{\;^1D_{1}}^{\prime}}{x_{1} x_{2}\left(M^{\prime 2}
	-M_{0}^{\prime 2}\right)}\left(\frac{m_{1}^{\prime}-m_{2}}{w_{\;^1D_{1}}^{\prime}} p_{\perp}^{\prime 2}\right),
\en
where $m_{1}^{\prime}$ and $m_{2}$ represent the constituent quarks of the states $D^*_0, \ ^3D_{1}$ and $^1D_{1}$. The decay constants can be obtained through experimental measurements for the purely leptonic decays or theoretical calculations. The explicit forms of $h^\prime_{M}$  are given by \cite{Cheng:2003sm}
\be
h_{D^*_0}^{\prime} &=&\sqrt{\frac{2}{3}} h_{\;^3D_{1}}^{\prime}=\left(M^{\prime 2}-M_{0}^{\prime 2}\right) \sqrt{\frac{x_{1} x_{2}}{N_{c}}} \frac{1}{\sqrt{2} \widetilde{M}_{0}^{\prime}}
\frac{\widetilde{M}_{0}^{\prime 2}}{2 \sqrt{3} M_{0}^{\prime}} \varphi_{p}^{\prime},\label{hs3a}\\
h_{\;^1D_{1}}^{\prime} &=&\left(M^{\prime2}-M_{0}^{\prime 2}\right) \sqrt{\frac{x_{1} x_{2}}{N_{c}}} \frac{1}{\sqrt{2} \widetilde{M}_{0}^{\prime}} \varphi_{p}^{\prime}.
\label{h1a}\en
\subsection{Helicity amplitudes and observables  }

 Since the form factors involving the fitted parameters for most of the transitions $B_{(s)} \to D_{(s)}^{**}$ have been investigated in our recent work \cite{Zhang:2023ypl}, so it is convenient to obtain the differential decay widths of these semileptontic $B$ decays by the combination of the helicity amplitudes via form factors, which are listed as following
 \begin{footnotesize}
\begin{eqnarray}
 \frac{d\Gamma(B \to D^*_0\ell\nu_\ell)}{dq^2} &=&(\frac{q^2-m_\ell^2}{q^2})^2\frac{ {\sqrt{\lambda(m_{B}^2,m_{D^*_0}^2,q^2)}} G_F^2 |V_{cb}|^2} {384m_{B}^3\pi^3}
 \times \frac{1}{q^2} \nonumber\\
 &&\;\;\;\times \left\{ (m_\ell^2+2q^2) \lambda(m_{B}^2,m_{D^*_0}^2,q^2) F_1^2(q^2)  +3 m_\ell^2(m_{B}^2-m_{D^*_0}^2)^2F_0^2(q^2)
\right\},\label{eq:pp}
\en
\be
 \frac{d\Gamma_L(B\to D^{(\prime)}_{1}\ell\nu_\ell)}{dq^2}&=&(\frac{q^2-m_\ell^2}{q^2})^2\frac{ {\sqrt{\lambda(m_{B}^2,m_{D^{(\prime)}_{1}}^2,q^2)}} G_F^2 |V_{cb}|^2} {384m_{B}^3\pi^3}
 \times \frac{1}{q^2} \left\{ 3 m_\ell^2 \lambda(m_{B}^2,m_{D^{(\prime)}_{1}}^2,q^2) V_0^2(q^2)+(m_\ell^2+2q^2)\right.\nonumber\\
 &&\times  \left. \left|\frac{1}{2m_{D^{(\prime)}_{1}}}  \left[
 (m_{B}^2-m_{D^{(\prime)}_{1}}^2-q^2)(m_{B}-m_{D^{(\prime)}_{1}})V_1(q^2)-\frac{\lambda(m_{B}^2,m_{D^{(\prime)}_{1}}^2,q^2)}{m_{B}-m_{D^{(\prime)}_{1}}}V_2(q^2)\right]\right|^2
\right\},\label{eq:decaywidthlon}\\
 \frac{d\Gamma_\pm(B\to
 {D^{(\prime)}_{1}}\ell\nu_\ell)}{dq^2}&=&(\frac{q^2-m_\ell^2}{q^2})^2\frac{ {\sqrt{\lambda(m_{B}^2,m_{D^{(\prime)}_{1}}^2,q^2)}} G_F^2 |V_{cb}|^2} {384m_{B}^3\pi^3}
  \nonumber\\
 &&\;\;\times \left\{ (m_\ell^2+2q^2) \lambda(m_{B}^2,m_{D^{(\prime)}_{1}}^2,q^2)\left|\frac{A(q^2)}{m_{B}-m_{D^{(\prime)}_{1}}}\mp
 \frac{(m_{B}-m_{D^{(\prime)}_{1}})V_1(q^2)}{\sqrt{\lambda(m_{B}^2,m_{D^{(\prime)}_{1}}^2,q^2)}}\right|^2
 \right\},
 \label{amptranse}
\end{eqnarray}
\end{footnotesize}
where $\lambda(a,b,c)=(a+b-c)^{2}-4ab$ and $m_{\ell}$ is the mass of the lepton $\ell$ with $\ell=e,\mu,\tau$\footnote{For now on, we will use $\ell$ to represent $e,\mu,\tau$ and use $\ell^\prime$ to represent $e,\mu$ for simplicity. }. The helicity amplitudes for the decays $B_s\to D^*_{s0}\ell\nu_\ell$ and $B_s\to D^{(\prime)}_{s1}\ell\nu_\ell$ can be obtained from Eqs. (\ref{eq:pp}), (\ref{eq:decaywidthlon}) and (\ref{amptranse}), respectively, with simple replacement.
The combined transverse and total differential decay widths are defined as
\be
\frac{d \Gamma_{T}}{d q^{2}}=\frac{d \Gamma_{+}}{d q^{2}}+\frac{d \Gamma_{-}}{d q^{2}}, \quad \frac{d \Gamma}{d q^{2}}=\frac{d \Gamma_{L}}{d q^{2}}+\frac{d \Gamma_{T}}{d q^{2}}.
\en

For the decays with $D^{(\prime)}_{1}$ and $D^{(\prime)}_{s1}$ involved, it is meaningful to define the polarization fraction due to the existence of different polarizations
\be
f_{L}=\frac{\Gamma_{L}}{\Gamma_{L}+\Gamma_{+}+\Gamma_{-}}. \label{eq:fl}
\en
As to the forward-backward asymmetry, the analytical expression is defined as \cite{ptau3},
\be
A_{FB} = \frac{\int^1_0 {d\Gamma \over dcos\theta} dcos\theta - \int^0_{-1} {d\Gamma \over dcos\theta} dcos\theta}
{\int^1_{-1} {d\Gamma \over dcos\theta} dcos\theta} = \frac{\int b_\theta(q^2) dq^2}{\Gamma_{B\to D^{**}\ell\nu_\ell}},\label{eq:AFB}
\en
where $\theta$ is the angle between the 3-momenta of the lepton $\ell$ and the initial $B$ meson in the $\ell\nu$ rest frame. The the angular coefficient $b_{\theta}(q^2)$ for the decays $B \to D^*_0\ell\nu_\ell$ is given as \cite{ptau3}
\be
b_\theta(q^2) = {G_F^2 |V_{cb}|^2 \over 128\pi^3 m_{B}^3} q^2 \sqrt{\lambda(q^2)}
\left( 1 - {m_\ell^2 \over q^2} \right)^2 {m_\ell^2 \over q^2} ( H^s_{V,0}H^s_{V,t} ) , \label{eq:btheta1}
\label{eq:btheta2}
\en
with the helicity amplitudes
\be
H^s_{V,0}\left(q^{2}\right)  =\sqrt{\frac{\lambda\left(q^{2}\right)}{q^{2}}} F_{1}\left(q^{2}\right),
H^s_{V,t}\left(q^{2}\right)  =\frac{m_{B}^{2}-m_{{D^*_0}}^{2}}{\sqrt{q^{2}}} F_{0}\left(q^{2}\right).
\en
Here $\lambda(q^2)=((m_B-m_{{D^*_0}})^2-q^2)((m_B+m_{{D^*_0}})^2-q^2)$. While for the decays $B\to D^{(\prime)}_{1}\ell\nu_\ell$, the function $b_{\theta}(q^2)$ is written as
\be
b_\theta(q^2) = {G_F^2 |V_{cb}|^2 \over 128\pi^3 m_{B}^3} q^2 \sqrt{\lambda(q^2)}
\left( 1 - {m_\ell^2 \over q^2} \right)^2 \left[ {1 \over 2}(H_{V,+}^2-H_{V,-}^2)+ {m_\ell^2 \over q^2} ( H_{V,0}H_{V,t} ) \right],
\en
where the corresponding helicity amplitudes are listed as 
\be
H_{V,\pm}\left(q^{2}\right)&=&\left(m_{B_{s}}-{m_{D_{1}}}\right) V_{1}\left(q^{2}\right) \mp \frac{\sqrt{\lambda\left(q^{2}\right)}}{m_{B}-m_{D_{1}}} A\left(q^{2}\right), \non
H_{V,0}\left(q^{2}\right)&=&\frac{m_{B}-m_{D_{1}}}{2 m_{D_{1}} \sqrt{q^{2}}}\left[-\left(m_{B}^{2}-m_{D_{1}}^{2}-q^{2}\right) V_{1}\left(q^{2}\right)+\frac{\lambda\left(q^{2}\right) V_{2}\left(q^{2}\right)}{\left(m_{B}-m_{D_{1}}\right)^{2}}\right],\non
H_{V,t}\left(q^{2}\right)&=&-\sqrt{\frac{\lambda\left(q^{2}\right)}{q^{2}}} V_{0}\left(q^{2}\right), \label{H}
\en
 It is noticed that the subscript $V$ in each helicity amplitude refers to the $\gamma_\mu(1-\gamma_5)$ current.

\section{Numerical results and discussions} \label{numer}
 \begin{table}[H]
\caption{The values of the input parameters \cite{BN,pdg22,Cheng:2003id,Becirevic:1998ua,Li:2009wq}.}
\label{tab:constant}
\begin{tabular*}{16.5cm}{@{\extracolsep{\fill}}l|cccc}
  \hline\hline
\textbf{Mass(\text{GeV})} &$m_{b}=4.8$
&$m_{c}=1.4$&$m_{u}=0.25$&$m_{s}=0.37$ \\[1ex]
&$m_{e}=0.000511$&$m_{\mu}=0.106$&$m_{\tau}=1.78$&$m_{B_s}=5.367$  \\[1ex]
&$m_{B}=5.279$ & $m_{D_{0}^{\ast}}=2.343$& $m_{D_{s0}^{\ast}}=2.317$&$m_{D_{s1}}=2.460$ \\[1ex]
&$m_{D_{s1}^{\prime}}=2.536$  & $m_{D_{1}}=2.422$  & $m_{D^{\prime}_1}=2.412 $\\[1ex]
\hline
\end{tabular*}
\begin{tabular*}{16.5cm}{@{\extracolsep{\fill}}l|l}
  \hline
{{\textbf{CKM}}} &$V_{cb}=(40.8\pm1.4) \times 10^{-3}$~~~~~~~~~~~~~~~~~~~~~~~~~~~~~~~~~~~~ \\[1ex]
\hline
\end{tabular*}
\begin{tabular*}{16.5cm}{@{\extracolsep{\fill}}l|ccc}
  \hline
{{\textbf{shape parameters{(GeV)}}}} &$\beta^{\prime}_{B}=0.555^{+0.048}_{-0.048}$&$\beta^{\prime}_{B_{s}}=0.628^{+0.035}_{-0.034}$&$\beta^{\prime}_{D_{0}^{\ast}}=0.373^{+0.063}_{-0.059}$\\[1ex]
&$\beta^{\prime}_{D_{s0}^{\ast}}=0.325^{+0.043}_{-0.043}$&$\beta^{\prime}_{^3D_{s1}}=0.342^{+0.030}_{-0.034}$
&$\beta^{\prime}_{^1D_{s1}^{\prime}}=0.342^{+0.039}_{-0.039}$\\[1ex]
&$\beta^{\prime}_{^3D_{1}}=0.332^{+0.031}_{-0.034}$&$\beta^{\prime}_{^1D_{1}^{\prime}}=0.329^{+0.038}_{-0.040}$\\[1ex]
\hline
\end{tabular*}
\begin{tabular*}{16.5cm}{@{\extracolsep{\fill}}l|cc}
	\hline
	{{\textbf{Lifetimes(s)}}} &$\tau_{B_s}=(1.520\pm0.005)\times 10^{-12}$&$\tau_{B^{\pm}}=(1.638\pm0.004)\times 10^{-12}$\\[1ex]
	&$\tau_{B^{0}}=(1.519\pm0.004)\times 10^{-12}$ \\[1ex]
	\hline\hline
\end{tabular*}\label{constants}
\end{table}
The adopted input parameters \cite{pdg22}, such as the constituent quark masses, the hadron and lepton masses, the $B$ meson lifetime and the Cabibbo-Kobayashi-Maskawa (CKM) matrix element $V_{cb}$, in our numerical calculations are listed in Table \ref{constants}.
In the calculations of the helicity amplitudes, the transition form factors are the most important inputs, some of which have been calculated in our previous work \cite{Zhang:2023ypl}.  The
parameterized form factors are extrapolated from the space-like region to the  time-like region  by using following expression,
\be
F\left(q^{2}\right)=\frac{F(0)}{1-a q^{2} / m^{2}+b q^{4} / m^{4}},\label{F}
\en
where $m$ represents the initial meson mass and $F(q^{2})$ denotes the different form factors.
The values of $a$ and $b$ can be obtained by performing a 3-parameter fit to the form factors in the range $-15 \text{GeV}^2\leq q^2\leq0$, which are collected in Table \ref{form factor1}. The uncertainties arise from the decay constants of the initial $B_{(s)}$ meson and the final state mesons. Certainly, in order to compare with the results given in other works, we also give the form factors of the transitions $B_{(s)}\to D^{3/2}_{(s)1}, D^{1/2}_{(s)1}$ under the heavy quark limit, which are shown in Table \ref{form D1}. Obviously, our results are consistent with the previous CLFQM \cite{Cheng:2003sm} and  ISGW2 \cite{Verma:2011yw} calculations. The signs of the form factors of the transitions $B_{(s)}\to D^{1/2}_{(s)1}$ between ours and the other two theoretical predictions \cite{Verma:2011yw, Cheng:2003sm} are opposite because the definations for the $D^{1/2}_{(s)1}$ mixing formula shown in Eq. (\ref{mixing2}) are different.

\begin{figure}[H]
	\vspace{0.5cm}
	\centering
	\subfigure[]{\includegraphics[width=0.45\textwidth]{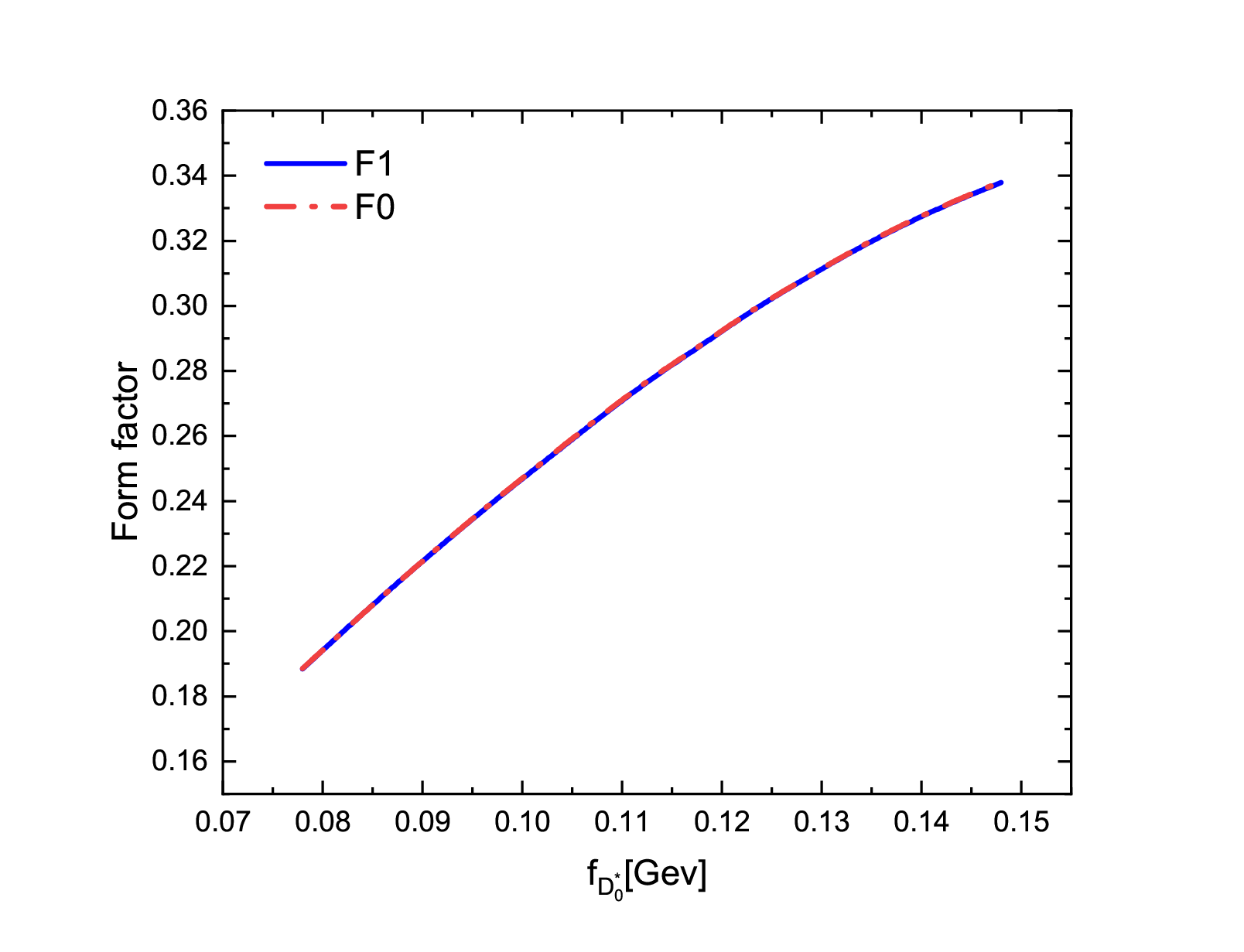}}
	\subfigure[]{\includegraphics[width=0.45\textwidth]{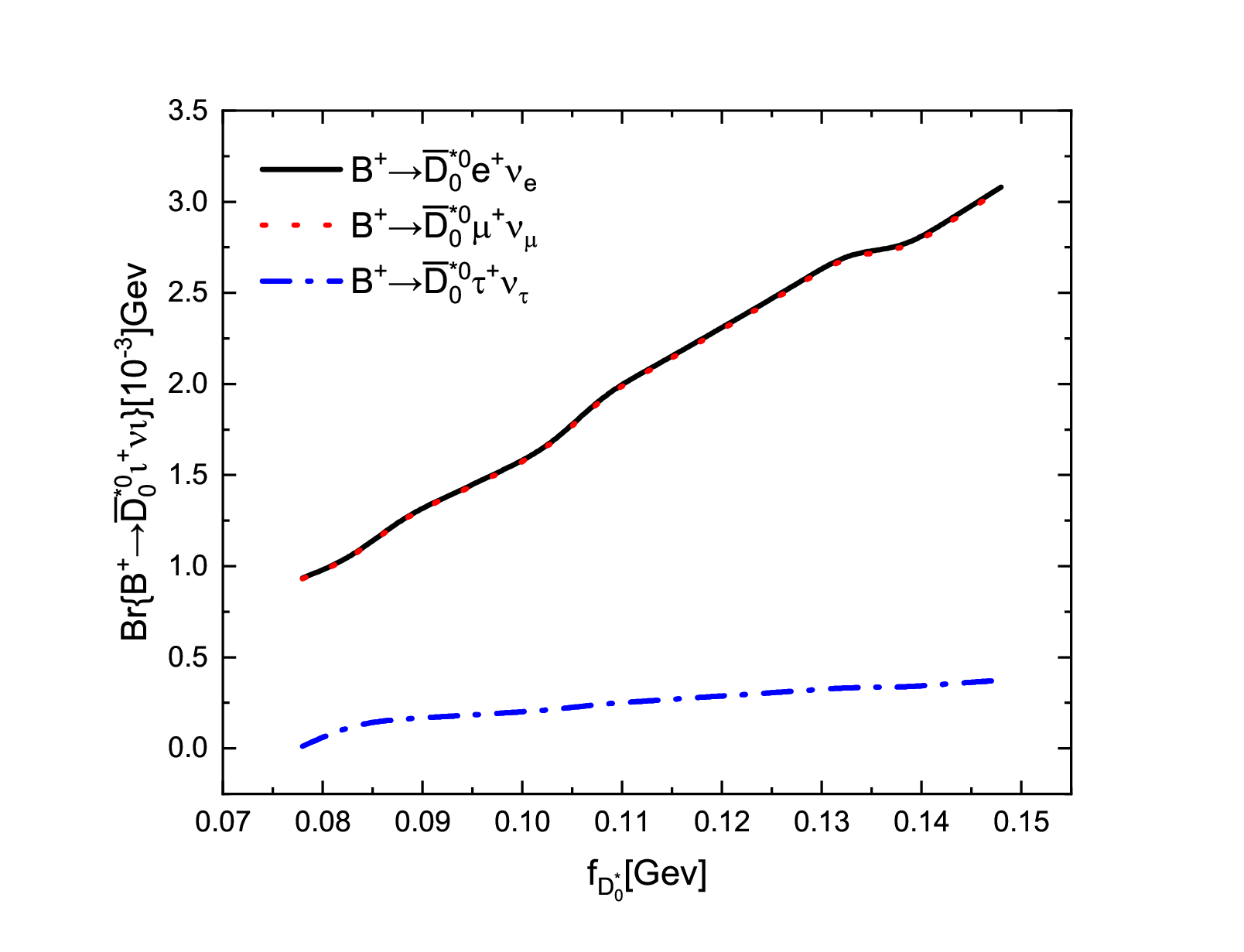}}
	\caption{The dependencies of the form factors of the transition $B\to D^*_{0}$ (left) and the branching ratios of the semileptonic decays $B\to D^*_{0}\ell\nu_{\ell}$ (right) on the decay constant $f_{D^*_{0}}$.}\label{figD0}
\end{figure}
\begin{table}[H]
	\caption{The form factors of the transtions $B_{(s)} \to D^*_0, D^*_{s0}, D_{s1}, D^\prime_{s1}, D_1 $ and $ D_1^\prime$ in the CLFQM. The uncertainties are from the decay constants of $B_{(s)}$ and the final state mesons.}
	\begin{center}
		\scalebox{0.9}{
			\begin{tabular}{ccccc}
				\hline\hline
				& $F_{i}(q^{2}=0)$&$F_{i}(q^{2}_{max})$&a&b\\
				\hline\hline
				$F_{1}^{BD^{\ast}_{0}}$&$0.25^{+0.03+0.05}_{-0.02-0.05}$&$0.30^{+0.03+0.06}_{-0.03-0.07}$&$0.70^{+0.04+0.03}_{-0.05-0.11}$&$0.65^{+0.08+0.03}_{-0.07-0.07}$\\
				$F_{0}^{BD^{\ast}_{0}}$&$0.25^{+0.03+0.05}_{-0.02-0.05}$&$0.22^{+0.02+0.04}_{-0.01-0.04}$&$-0.38^{+0.04+0.05}_{-0.04-0.02}$&$0.21^{+0.07+0.08}_{-0.07-0.08}$\\
				\hline
				$F_{1}^{B_s D^*_{s0}}$&$0.21^{+0.02+0.04}_{-0.01-0.04}$&$0.24^{+0.02+0.05}_{-0.01-0.05}$&$0.63^{+0.05+0.07}_{-0.06-0.12}$&$0.78^{+0.08+0.01}_{-0.09-0.04}$\\
				$F_{0}^{B_s D^*_{s0}}$&$0.21^{+0.02+0.04}_{-0.01-0.04}$&$0.18^{+0.02+0.03}_{-0.01-0.03}$&$-0.43^{+0.01+0.01}_{-0.00-0.02}$&$0.28^{+0.03+0.01}_{-0.06-0.04}$\\
				\hline
			$A^{ B_s D_{s1}}$&$0.20^{+0.01+0.02}_{-0.01-0.02}$&$0.18^{+0.02+0.03}_{-0.01-0.02} $&$-0.27^{+0.06+0.8}_{-0.07-0.09}$ & $0.11^{+0.02+0.02}_{-0.02-0.03}$\\
			$V^{ B_s D_{s1}}_{0}$&$0.40^{+0.02+0.04}_{-0.02-0.04}$&$0.42^{+0.02+0.05}_{-0.02-0.05} $&$-0.17^{+0.02+0.04}_{-0.04-0.06}$ & $-0.02^{+0.01+0.00}_{-0.00-0.01}$\\
			$V^{ B_s D_{s1}}_{1}$&$0.58^{+0.01+0.03}_{-0.02-0.04}$&$0.57^{+0.01+0.03}_{-0.02-0.04} $&$-0.05^{+0.01+0.01}_{-0.01-0.01}$ & $0.02^{+0.00+0.01}_{-0.00-0.00}$\\
			$V^{ B_s D_{s1}}_{2}$&$-0.05^{+0.01+0.02}_{-0.00-0.01}$&$ -0.05^{+0.01+0.03}_{-0.00-0.02}$&$0.56^{+0.06+0.22}_{-0.06-0.25}$ & $2.50^{+0.25+1.67}_{-0.20-1.30}$\\
			\hline
			$A^{ B_s D^{\prime}_{s1}}$&$0.08^{+0.01+0.02}_{-0.01-0.02}$&$0.03^{+0.01+0.01}_{-0.02-0.03}$&$2.05^{+0.13+0.35}_{-0.10-0.35}$ & $5.57^{+0.25+0.50}_{-0.20-0.41}$\\
			$V^{B_s D^{\prime}_{s1}}_{0}$&$-0.08^{+0.01+0.04}_{-0.01-0.04}$&$-0.05^{+0.02+0.05}_{-0.01-0.04}$&$1.24^{+0.05+0.23}_{-0.06-0.25}$ & $0.74^{+0.02+0.21}_{-0.02-0.17}$\\
			$V^{ B_s D^{\prime}_{s1}}_{1}$&$0.17^{+0.02+0.04}_{-0.03-0.03}$&$0.15^{+0.01+0.02}_{-0.03-0.03}$&$-0.52^{+0.06+0.06}_{-0.05-0.06}$ & $0.36^{+0.01+0.03}_{-0.00-0.08}$\\
			$V^{ B_s D^{\prime}_{s1}}_{2}$&$0.11^{+0.01+0.01}_{-0.02-0.02}$&$0.10^{+0.01+0.01}_{-0.02-0.02}$&$0.25^{+0.06+0.06}_{-0.07-0.07}$ & $-0.07^{+0.03+0.01}_{-0.04-0.03}$\\
				\hline
			$A^{ B D_{1}}$&$0.21^{+0.02+0.02}_{-0.01-0.01}$&$0.20^{+0.02+0.02}_{-0.02-0.02} $&$-0.23^{+0.10+0.09}_{-0.09-0.09}$ & $0.09^{+0.02+0.02}_{-0.02-0.02}$\\
			$V^{ B D_{1}}_{0}$&$0.42^{+0.03+0.04}_{-0.02-0.04}$&$ 0.44^{+0.04+0.05}_{-0.02-0.05}$&$0.16^{+0.02+0.03}_{-0.02-0.05}$ & $-0.02^{+0.00+0.01}_{-0.00-0.01}$\\
			$V^{ B D_{1}}_{1}$&$0.57^{+0.03+0.03}_{-0.01-0.04}$&$ 0.56^{+0.03+0.03}_{-0.02-0.04}$&$-0.07^{+0.00+0.01}_{-0.01-0.00}$ & $0.02^{+0.00+0.00}_{-0.00-0.00}$\\
			$V^{ B D_{1}}_{2}$&$-0.06^{+0.01+0.02}_{-0.01-0.02}$&$ -0.06^{+0.01+0.04}_{-0.01-0.03}$&$0.57^{+0.06+0.17}_{-0.07-0.20}$ & $2.13^{+0.21+1.30}_{-0.18-1.17}$\\
			\hline
			$A^{ B D^{\prime}_{1}}$&$0.07^{+0.02+0.02}_{-0.02-0.02}$&$ 0.02^{+0.03
			+0.01}_{-0.02-0.02}$&$-1.90^{+0.09+0.12}_{-0.10-0.14}$ & $5.64^{+0.22+1.59}_{-0.18-1.29}$\\
			$V^{B D^{\prime}_{1}}_{0}$&$-0.08^{+0.02+0.04}_{-0.02-0.03}$&$-0.06^{+0.02+0.06}_{-0.02-0.03} $&$-0.11^{+0.01+0.06}_{-0.04-0.11}$ & $3.67^{+0.05+0.51}_{-0.05-0.51}$\\
			$V^{ B D^{\prime}_{1}}_{1}$&$0.14^{+0.04+0.03}_{-0.04-0.03}$&$ 0.13^{+0.03+0.03}_{-0.04-0.02}$&$-0.36^{+0.04+0.04}_{-0.06-0.06}$ & $0.09^{+0.01+0.01}_{-0.01-0.01}$\\
			$V^{ B D^{\prime}_{1}}_{2}$&$0.10^{+0.03+0.02}_{-0.02-0.02}$&$0.11^{+0.03+0.02}_{-0.04-0.04} $&$0.19^{+0.06+0.06}_{-0.09-0.09}$ & $0.00^{+0.01+0.01}_{-0.00-0.01}$\\
				\hline\hline
			\end{tabular}\label{form factor1}
		}
	\end{center}
\end{table}

The branching ratios of the decays $B_{(s)} \to D^*_{(s)0} \ell {\nu}_{\ell}$ are shown in Table \ref{tab2}. For comparison, the values given by other theoretical approaches and the current experiments are also listed. One can find that the branching ratios of the decays $B^+\to \bar{D}^{*0}_0 \ell^+ {\nu}_{\ell}$ are larger than the results from the QCD sum rules (QCDSRs) \cite{Zuo:2023ksq} and the heavy quark effective field theory (HQEFT) \cite{W.Y.}. Obviously, our predictions are consistent with the previous CLFQM calculations \cite{Kang:2018jzg} and the differences 
are mainly due to the different values of the decay constant $f_{D^*_0}$. In Figure \ref{figD0}, the changing trends of the form factors $F^{BD^*_0}_1(q^2=0)  (F^{BD^*_0}_0(q^2=0))$ and the branching ratios of the decays $B^+\to \bar D^{*0}_0 \ell^+\nu_\ell$ with $f_{D^*_0}$ are plotted, respectively. Both of them increase with $f_{D^*_0}$. The branching ratios of the decays $B^0\to D^{*-}_0 \ell^+ {\nu}_{\ell}$ can agree with QCD LCSRs under scenario 2 (S2), where the broad resonance $D^*_0$ was considered as consisting of the two resonances $D^*_0(2105)$ and $D^*_0(2451)$ \cite{Gubernari:2023rfu}.  While it seems to be too large in scenario 1 (S1), where $D^*_0$ was considered as a single resonance \cite{Gubernari:2023rfu}. Certainly, there still exist large errors in both S1 and S2. Our predictions are much smaller than those given in the LCSRs approach \cite{Shen:2012mm}, where a very large form factor $F_0^{BD^*_0}(q^2=0)=0.94$ was used in the calculations. There exists a similar situation for the decays $B_{s}^{0}\to D^{*-}_{s0} \ell^+ {\nu}_{\ell}$. It is surprising that a large value $(2.2\pm0.86)\times 10^{-2}$ was measured by Belle \cite{Belle:2007uwr} in 2008, latter a more large one $(4.4\pm1.0)\times10^{-2}$ was given by BaBar \cite{BaBar:2008ozy}, while Belle updated their measurement with only a small upper limit $<0.44\times10^{-3}$ obtained. In theory, the branching ratios of the decays $B^{+}\to \bar{D}^{*0}_0 \ell^{\prime+}\nu_{\ell^\prime}$ and $B^{0}\to D^{*-}_0 \ell^{\prime+}\nu_{\ell^\prime}$ should be not much difference. In order to clarifying this puzzle, we urge our experimental colleagues to perform further more precise measurements.  In Ref. \cite{A.L.}, the authors calculated the branching ratios based on the general HQET expansion, the so called the Leibovich-Ligeti-Stewart-Wise (LLSW) scheme, combining other theoretical results and the constrains from experimental measurements, which are smaller than nearly all the present avaiable predictions. 
\begin{table}[H]
	\caption{Form factors of the transitions $B_{(s)} \to D^{3/2}_{s1},D^{1/2}_{s1},D^{3/2}_{1}$ and $D^{1/2}_{1}$ at $q^{2}= 0$ together with other theoretical results. }
	\begin{center}
		\scalebox{1.0}{
			\begin{tabular}{c|c|cccc}
				\hline\hline
				Transitions  &References&$A(0)$&$V_{0}(0)$&$V_{1}(0)$&$V_{2}(0)$\\
				\hline\hline
				$B_{s}\to D^{3/2}_{s1} $&This work&$0.19$&$0.41$&$0.55$&$-0.07$\\
				&CLFQM$^a$ \cite{Verma:2011yw}&$0.24$&$0.49$&$0.57$&$-0.09$\\
				\hline
				$B_{s}\to D^{1/2}_{s1} $&This work&$0.10$&$-0.03$&$0.24$&$0.11$\\
				&CLFQM$^a$ \cite{Verma:2011yw}&$-0.17$&$0.13$&$-0.25$&$-0.17$\\
				\hline
				$B\to D^{3/2}_{1} $&This work&$0.20$&$0.43$&$0.55$&$-0.07$\\
				\hline
				&CLFQM \cite{Cheng:2003sm}&$0.23$&$0.47$&$0.55$&$-0.09$\\
				&ISGW2 \cite{Cheng:2003sm}&$0.16$&$0.43$&$0.40$&$-0.12$\\
				&CLFQM \cite{Verma:2011yw}&$0.25$&$0.52$&$0.58$&$-0.10$\\
				\hline
				$B\to D^{1/2}_{1} $&This work&$0.09$&$-0.02$&$0.21$&$0.09$\\
				\hline
				&CLFQM$^a$ \cite{Cheng:2003sm}&$-0.12$&$0.08$&$-0.19$&$-0.12$\\
				&ISGW2$^a$\cite{Cheng:2003sm}&$-0.16$&$0.18$&$-0.19$&$-0.18$\\
				&CLFQM$^a$ \cite{Verma:2011yw}&$-0.13$&$0.11$&$-0.19$&$-0.14$\\
				\hline\hline
			\end{tabular}\label{form D1}
		}
	\end{center}
	{\footnotesize $^a$ Due to the signs of the $D^{1/2}_{(s)1}$ mixing formula Eq. (\ref{mixing2}) between this work and Refs. \cite{Cheng:2003sm,Verma:2011yw} are opposite, the corresponding results are just contrary with our predictions.} \\
\end{table}
\begin{table}[H]
	\caption{Branching ratios ($10^{-3}$) of the semileptonic decays $B_{(s)}\to D^*_{(s)0}\ell\nu_\ell$.}
	\begin{center}
		\scalebox{0.72}{
			\begin{tabular}{c|c|c|c}
				\hline\hline
				References&$B^{+}\to \bar{D}^{*0}_0 e^{+}\nu_{e}$&$B^{+}\to \bar{D}^{*0}_0 \mu^{+}\nu_{\mu}$&$B^{+}\to \bar{D}^{*0}_0 \tau^{+}\nu_{\tau}$\\
				\hline\hline
				This work&$1.66^{+0.43+0.74}_{-0.27-0.62}$&$1.65^{+0.43+0.74}_{-0.26-0.62}$&$0.21^{+0.06+0.10}_{-0.03-0.08}$\\
				\hline
			 QCDSRs	\cite{Zuo:2023ksq}&$0.61^{+0.28+0.04}_{-0.21-0.05}$&$0.61^{+0.28+0.04}_{-0.21-0.05}$&$0.04^{+0.02+0.00}_{-0.01-0.00}$\\
				CLFQM \cite{Kang:2018jzg} &$2.31\pm0.25$&$2.31\pm0.25$&$0.30\pm0.03$\\
				HQEFT \cite{W.Y.} &$0.50\pm0.16$&$0.50\pm0.16$&$-$\\
                PDG \cite{pdg22}&$1.35\pm0.75$&$1.35\pm0.75$&$-$\\
				\hline
				References&$B^{0}\to D^{*-}_0 e^{+}\nu_{e}$&$B^{0}\to D^{*-}_0 \mu^{+}\nu_{\mu}$&$B^{0}\to D^{*-}_0 \tau^{+}\nu_{\tau}$\\
				\hline
					This work&$1.54^{+0.40+0.69}_{-0.25-0.57}$&$1.53^{+0.40+0.69}_{-0.25-0.57}$&$0.19^{+0.05+0.09}_{-0.03-0.07}$\\
				\hline
				QCD LCSRs \cite{Gubernari:2023rfu}$^a$ &$3.6^{+5.1}_{-3.0}$&$3.6^{+5.1}_{-3.0}$&$0.39^{+0.51}_{-0.31}$\\
				QCD LCSRs \cite{Gubernari:2023rfu}$^b$&$1.6^{+3.2}_{-1.4}$&$1.6^{+3.2}_{-1.4}$&$0.24^{+0.47}_{-0.21}$\\
                LCSRs \cite{Shen:2012mm} &$8.7^{+5.1}_{-2.8}$&$8.7^{+5.1}_{-2.8}$&$1.1^{+0.6}_{-0.3}$\\
				LLSW \cite{A.L.}&$0.51\pm0.12$&$0.51\pm0.12$&$0.050\pm0.013$\\
				BaBar \cite{BaBar:2008ozy} &$4.4\pm1.0$&$4.4\pm1.0$&$-$\\
				Belle \cite{Belle:2007uwr} &$2.0\pm0.86$&$2.0\pm0.86$&$-$\\
                    Belle \cite{Belle2023} &$<0.44$&$<0.44$&$-$\\
				\hline
				References&$B_{s}^{0}\to D^{*-}_{s0} e^{+}\nu_{e}$&$B_{s}^{0}\to D^{*-}_{s0} \mu^{+}\nu_{\mu}$&$B_{s}^{0}\to D^{*-}_{s0} \tau^{+}\nu_{\tau}$\\
				\hline
				This work&$1.26^{+0.26+0.55}_{-0.13-0.45}$&$1.25^{+0.26+0.55}_{-0.13-0.45}$&$0.18^{+0.04+0.08}_{-0.02-0.07}$\\
				\hline
				QCDSRs \cite{Zuo:2023ksq}&$0.72^{+0.30+0.04}_{-0.26-0.06}$&$0.71^{+0.33+0.04}_{-0.25-0.05}$&$0.06^{+0.03+0.00}_{-0.02-0.00}$\\
				CUM \cite{Navarra:2015iea}&$1.3$&$1.3$&$-$\\
				QCD	LCSRs \cite{Gubernari:2023rfu}&$1.9^{+3.8}_{-1.7}$&$1.9^{+3.8}_{-1.7}$&$0.26^{+0.49}_{-0.22}$\\
				QCDSRs \cite{M. Q.} &$0.9\sim2.0$&$0.9\sim2.0$&$-$\\
				QCDSRs \cite{T. M.} &$1.0$&$1.0$&$0.1$\\
				LCSRs \cite{Li:2009wq}&$2.3^{+1.2}_{-1.0}$&$2.3^{+1.2}_{-1.0}$&$0.57^{+0.28}_{-0.23}$\\
				CQM \cite{Zhao:2006at} &$4.90\sim5.71$&$4.9\sim5.71$&$-$\\
				RQM \cite{Faustov:2012mt}&$3.6\pm0.4$&$3.6\pm0.4$&$0.19\pm0.02$\\
				LSCRs \cite{Shen:2012mm}&$6.0\pm1.9$&$6.0\pm1.9$&$0.82^{+0.18}_{-0.20}$\\
				\hline
				\hline
			\end{tabular}\label{tab2}
		}
	\end{center}
	{\footnotesize $^a$ Results obtained in scenario 1 (S1), where $D^*_0$ was considered as a single broad resonance with mass being $(2343\pm10)$ MeV and width $(229\pm16)$ MeV. }\\
	{\footnotesize $^b$ Results obtained in scenario 2 (S2), where $D^*_0$ was assumed to consist of two scalar resonances $D^*_0(2105)$ and $D^*_0(2451)$.}\\
\end{table}

Then we compare our predictions for the branching ratios of the decays $B_{s}^{0}\to D^{*-}_{s0} \ell^+ {\nu}_{\ell}$ with the results obtained from other approaches. One can find that our results are consistent with those given in the chiral unitary approach (CUA) \cite{Navarra:2015iea}, the QCDSRs \cite{Zuo:2023ksq}, the QCD LCSRs \cite{Gubernari:2023rfu} and the LCSRs \cite{Li:2009wq} within errors. While they are much smaller than the constituent quark meson (CQM) \cite{Zhao:2006at} and the LCSRs \cite{Shen:2012mm} calculations. Although the LCSRs was used in both Ref. \cite{Li:2009wq} and Ref. \cite{Shen:2012mm}, their results are very different. It is because of the different correlation function, which is taken between the vacuum and $D^*_{s0} (B)$ with the $B (D^*_{s0})$ meson being interploated by a local current for the former (the latter). The form factor of the transition $B_s\to D^*_{s0}$ obtained in the latter (the so called B-meson LCSRs) is about $0.80$, which is larger than $0.53$ calculated by the former (the so called the conventional light meson LCSRs). 
Further experimental and theoretical researches are needed to clarify these divergences and puzzles.

\begin{table}[H]
	\caption{Branching ratios ($10^{-3}$) of the semileptonic decays $B\to D^{(\prime)}_{1}\ell\nu_\ell$ and $B_{s}\to D^{(\prime)}_{s1}\ell\nu_\ell$.}
	\begin{center}
		\scalebox{0.7}{
			\begin{tabular}{c|c|c|c}
				\hline\hline
                References&$B^{+}\to \bar{D}_1^{0} e^{+}\nu_{e}$&$B^{+}\to \bar{D}_1^{0} \mu^{+}\nu_{\mu}$&$B^{+}\to \bar{D}_1^{0}\tau^{+}\nu_{\tau}$\\
						\hline
					This work&$6.21^{+0.94+0.43+1.00}_{-0.67-0.28-0.90}$&$6.16^{+0.93+0.42+0.99}_{-0.66-0.27-0.89}$&$0.57^{+0.08+0.03+0.07}_{-0.05-0.04-0.07}$\\
					\hline
					QCDSRs \cite{Zuo:2023ksq} &$7.26^{+3.60+0.51}_{-2.87-0.49}$&$7.19^{+3.56+0.50}_{-2.84-0.48}$&$0.46^{+0.22+0.03}_{-0.17-0.03}$\\
					LLSW \cite{A.L.}&$6.40\pm0.44$&$6.40\pm0.44$&$0.63\pm0.06$\\
				HQEFT \cite{W.Y.2}&$5.9\pm1.6$&$5.9\pm1.6$&$-$\\
                PDG \cite{pdg22} &$4.26\pm0.26$&$4.26\pm0.26$&$-$\\                
					\hline
					References &$B^{+}\to \bar{D}_1^{\prime0} e^{+}\nu_{e}$&$B^{+}\to \bar{D}_1^{\prime0} \mu^{+}\nu_{\mu}$&$B^{+}\to \bar{D}_1^{\prime0}\tau^{+}\nu_{\tau}$\\
						\hline
					This work&$0.15^{+0.15+0.07+0.08}_{-0.09-0.03-0.06}$&$0.15^{+0.15+0.07+0.08}_{-0.09-0.03-0.06}$&$0.02^{+0.01+0.00+0.00}_{-0.01-0.02-0.00}$\\
					\hline
					QCDSRs \cite{Zuo:2023ksq} &$0.68^{+0.31+0.04}_{-0.24-0.05}$&$0.67^{+0.31+0.04}_{-0.23-0.05}$&$0.05^{+0.05+0.00}_{-0.02-0.00}$\\
					LLSW \cite{A.L.}&$0.46\pm0.37$&$0.46\pm0.37$&$0.03\pm0.03$\\
					HQEFT \cite{W.Y.}&$0.48\pm0.16$&$0.48\pm0.16$&$-$\\
                        PDG \cite{pdg22}&$2.55\pm0.90$&$2.55\pm0.90$&$-$\\
                \hline
				References&$B_{s}^{0}\to D_{s1}^- e^{+}\nu_{e}$&$B_{s}^{0}\to D_{s1}^- \mu^{+}\nu_{\mu}$&$B_{s}^{0}\to D_{s1}^- \tau^{+}\nu_{\tau}$\\
				\hline
				This work&$6.17^{+0.94+0.21+0.42}_{-0.91-0.62-0.52}$&$6.13^{+0.91+0.21+0.42}_{-0.94-0.61-0.52}$&$0.63^{+0.03+0.04+0.00}_{-0.05-0.06-0.00}$\\
				\hline
			QCDSRs	\cite{Zuo:2023ksq} &$6.31^{+3.07+0.44}_{-2.44-0.43}$&$6.25^{+3.03+0.44}_{-2.42-0.42}$&$0.38^{+0.18+0.03}_{-0.14-0.03}$\\
				CQM \cite{Zhao:2006at} &$7.52\sim8.69$&$7.52\sim8.69$&$-$\\
				QCDSRs \cite{T.M.2} &$4.90$&$4.90$&$-$\\
				RQM \cite{Faustov:2012mt} &$8.40\pm0.90$&$8.40\pm0.90$&$0.49\pm0.05$\\
				\hline
				References&$B_{s}^{0}\to D_{s1}^{\prime-} e^{+}\nu_{e}$&$B_{s}^{0}\to D_{s1}^{\prime-} \mu^{+}\nu_{\mu}$&$B_{s}^{0}\to D_{s1}^{\prime-} \tau^{+}\nu_{\tau}$\\
				\hline
				This work&$0.18^{+0.06+0.07+0.07}_{-0.07-0.07-0.09}$&$0.18^{+0.06+0.07+0.07}_{-0.07-0.07-0.09}$&$0.02^{+0.01+0.01+0.00}_{-0.00-0.01-0.00}$\\
				\hline
				QCDSRs \cite{Zuo:2023ksq}&$0.65^{+0.30+0.04}_{-0.23-0.05}$&$0.64^{+0.30+0.04}_{-0.23-0.05}$&$0.05^{+0.03+0.00}_{-0.02-0.00}$\\
			RQM	\cite{Faustov:2012mt} &$1.90\pm0.02$&$1.90\pm0.02$&$0.15\pm0.02$\\
				\hline\hline
			\end{tabular}\label{tab3}
		}
	\end{center}
\end{table}

\begin{figure}[H]
	\vspace{0.4cm}
	\centering
    \subfigure[]{\includegraphics[width=0.6\textwidth]{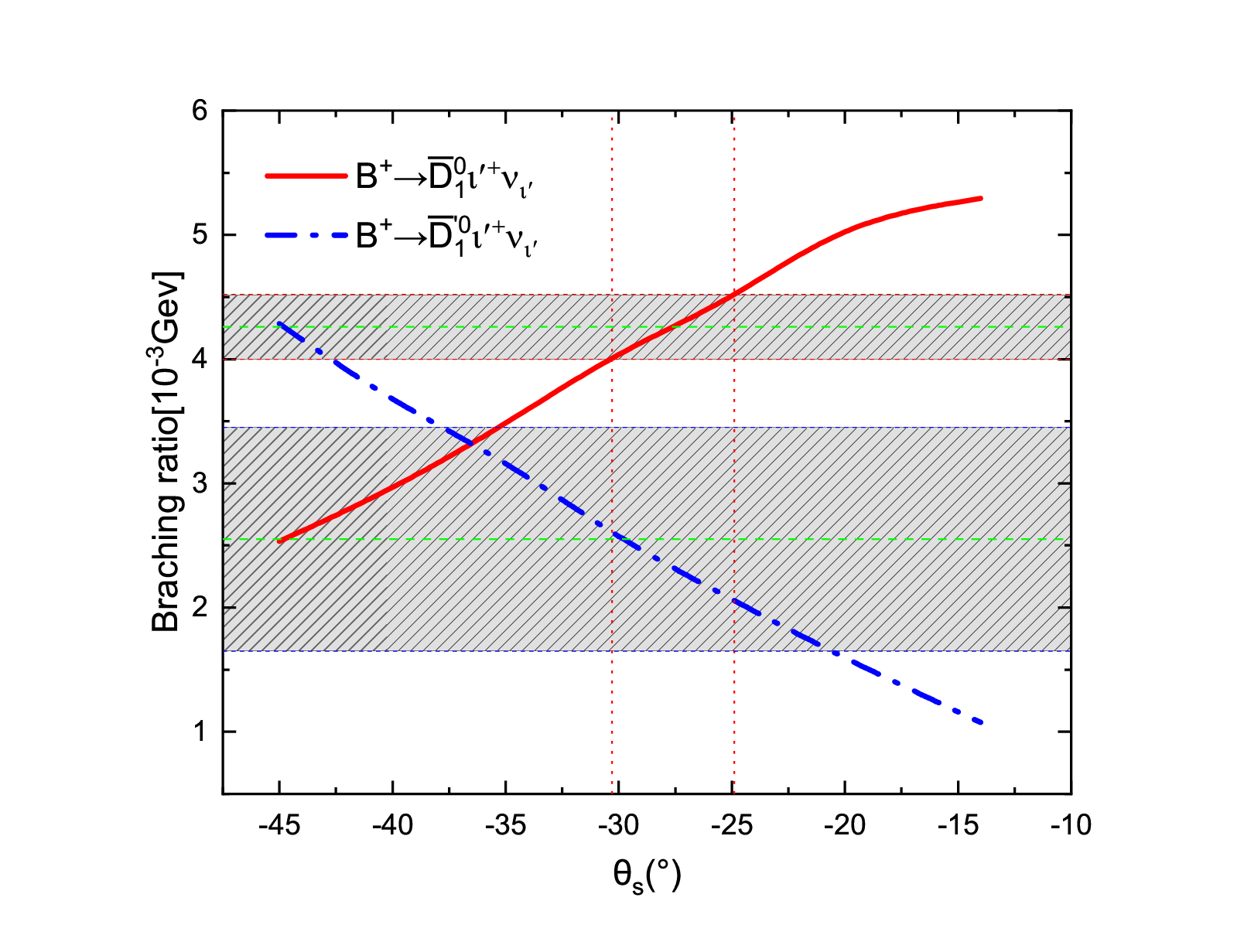}} \\   
	\caption{The mixing angle $\theta_s$ dependencies of the branching ratios for the semileptonic decays  $B^+\to \bar{D}^{0}_{1}\ell^{\prime+} \nu_{\ell^\prime}$ (the red solid line) and $B^+\to \bar{D}^{\prime0}_{1}\ell^{\prime+} \nu_{\ell^\prime}$ (the blue dash-dotted line). The upper (lower) shadow band and its horizontal image center line refers to the experimentally achievable range and the center value for the branching ratios of the decays $B^+\to D^0_1 \ell^{\prime}\nu_{\ell^\prime} (B^+\to \bar D^{\prime0}_1 \ell^{\prime}\nu_{\ell^\prime})$, respectively. }\label{figtheta}
    \end{figure}

\begin{figure}[H]
	\vspace{0.50cm}
	\centering
    \subfigure[]{\includegraphics[width=0.40\textwidth]{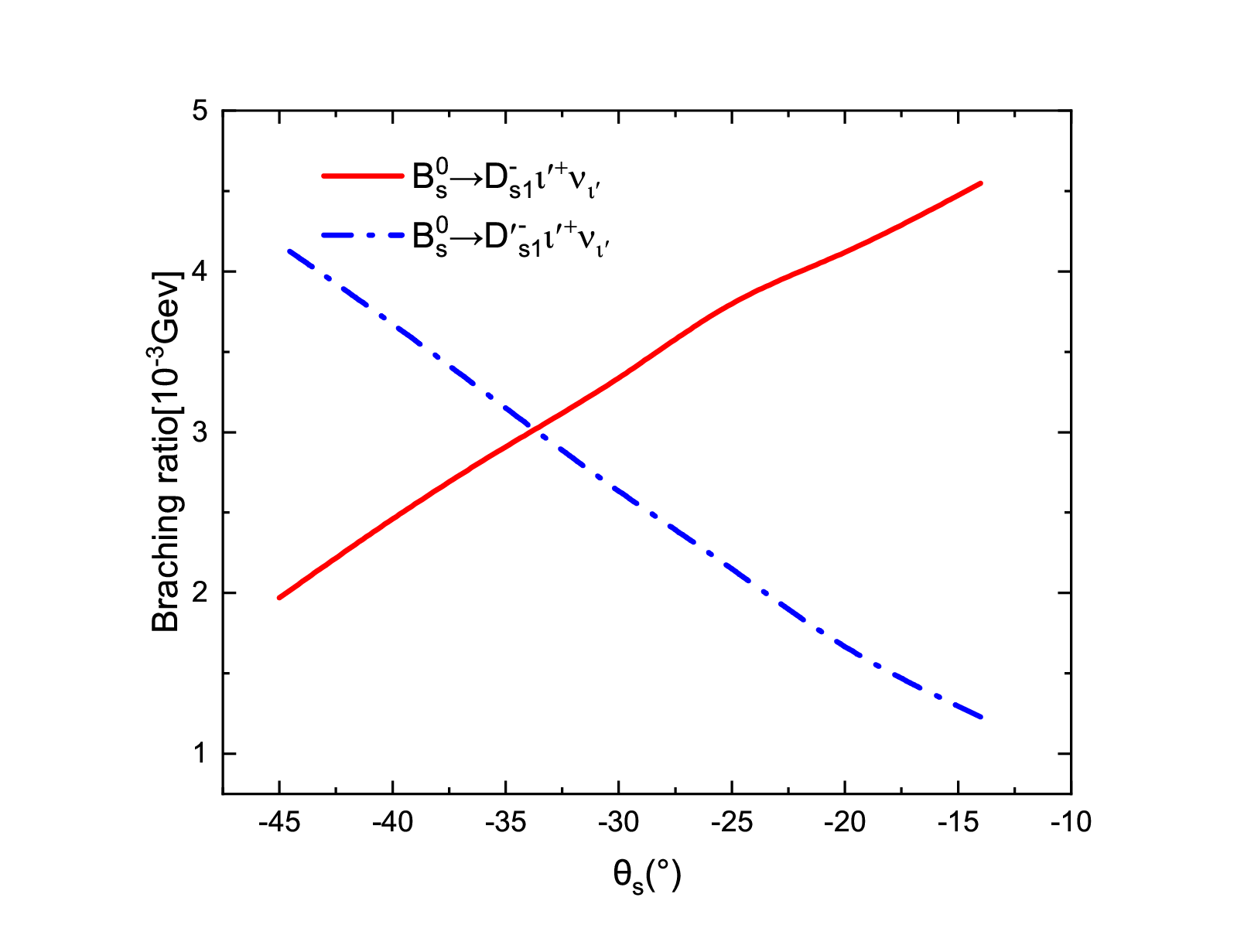}}
    \subfigure[]{\includegraphics[width=0.40\textwidth]{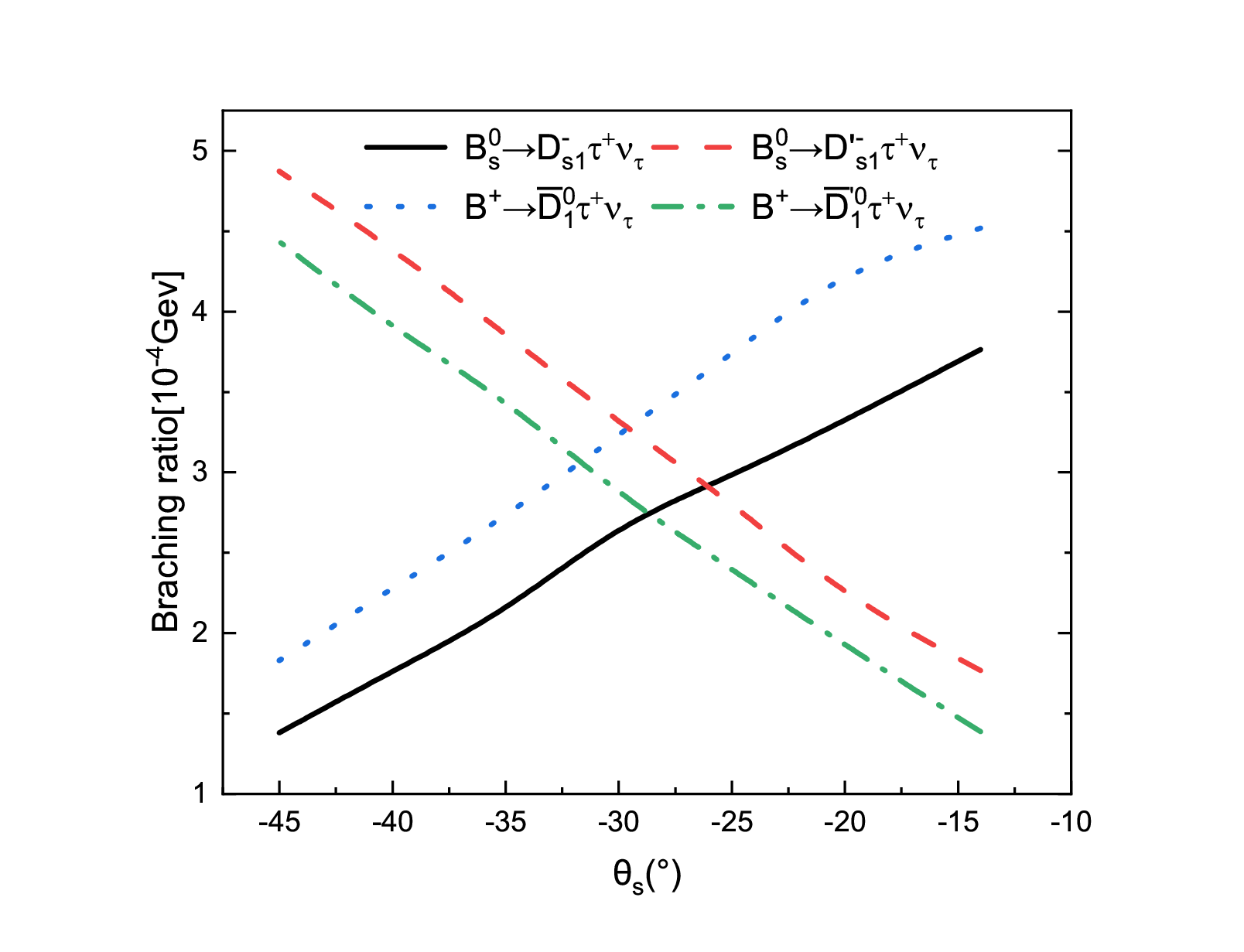}}\\
	\caption{The mixing angle $\theta_s$ dependencies of the branching ratios for the semileptonic decays  $B^0_s\to {D}^{(\prime)-}_{s1}\ell^{+} \nu_{\ell}$ (a) and $B^+\to \bar{D}^{(\prime)0}_{1}\tau^{+} \nu_{\tau}, B^0_s\to {D}^{(\prime)-}_{s1}\tau^{+} \nu_{\tau}$ (b). }\label{figtheta2}   
\end{figure}

We calculate the branching ratios of the decays $B^0_s\to D^{(\prime)-}_{s1}\ell^+ \nu_{\ell}$ and $B^+\to \bar{D}^{(\prime)0}_{1}\ell^+ \nu_{\ell}$, which are listed in Table \ref{tab3} with other theoretical predictions and data for comparison.  All the theoretical predictions show that the branching ratios of the decays $B^0_s\to D^-_{s1}\ell^+ {\nu}_{\ell}$ are (much) lager than those of the decays $B^0_s\to D^{\prime-}_{s1}\ell^+ \nu_{\ell}$. This is because that the related form factors of the transition $B_s\to D_{s1}$ are much larger than those of the transition $B_s\to D^{\prime}_{s1}$. There exists a similar situation between the decays $B^+\to \bar{D}^{0}_{1}\ell^+ \nu_{\ell}$ and $B^+\to \bar{D}^{\prime0}_{1}\ell^+ \nu_{\ell}$.
One can find that the branching ratios of the decays $B_{(s)}\to D_{(s)1}\ell \nu_{\ell}$ are comparable with the results given by most theoretical calculations, such as the QCDSRs \cite{Zuo:2023ksq, T.M.2}, the CQM \cite{Zhao:2006at}, the relativistic quark model (RQM) \cite{Faustov:2012mt}, the LLSW \cite{A.L.}, the HQEFT \cite{W.Y.}, and so on. Certainly, they are also consistent well with the present avalable data \cite{pdg22}. While for the decays $B^+\to \bar{D}^{\prime0}_{1}\ell^+ \nu_{\ell}$, their branching ratios given by all the theoretical predictions are smaller than the data measured by Belle \cite{Belle:2022yzd}.   
Therefore we urge our experimental colleagues to accurately measure these decays. It is very helpful to probe the
inner structures of the resonant states $D_{(s)1}$ and $D^{\prime}_{(s)1}$ by clarifying the tension between theory and experiment, which is the so called ‘1/2 vs 3/2 puzzle’. In order to explain this puzzle, we calculate the dependences of the branching ratios of the decays $B\to D^{(\prime)}_1 \ell^{\prime}\nu_{\ell^\prime}$ on the mixing angle $\theta_s$, which are shown in Figure \ref{figtheta}, where the branching ratios for the decays $B^+\to D^{(\prime)0}_1 \ell^{\prime}\nu_{\ell^\prime}$ increase (decrease) with the mixing angle $\theta_s$. The upper (lower) shadow band and its horizontal image center line refers to the experimentally achievable range and the center value for the branching ratios of the decays $B^+\to D^0_1 \ell^{\prime}\nu_{\ell^\prime} (B^+\to \bar D^{\prime0}_1 \ell^{\prime}\nu_{\ell^\prime})$, respectively. One can find that taking some negative mixing angle $\theta_s$ values within a range from $-30.3^\circ$ to $-24.9^\circ$ can explain the data, which correspond to $\theta$ within the range $5^\circ\sim10.4^\circ$.  It is similar for the decays $B^0_s\to {D}^{(\prime)-}_{s1}\ell^{+} \nu_{\ell}$ and $B^+\to \bar{D}^{(\prime)0}_{1}\tau^{+} \nu_{\tau}, B^0_s\to {D}^{(\prime)-}_{s1}\tau^{+} \nu_{\tau}$, the dependencies of their branching ratios on the mixing angle $\theta_s$ are shown in Figure \ref{figtheta2}. All of these decays shown that the branching ratios of the decays with $D_{(s)1}$ involved increase with the mixing angle $\theta_s$, while it is contrary for those of the decays  with  $D^\prime_{(s)1}$ involved.
    
In Table \ref{R},  we also calculate the lepton flavor universality ratios, which are defined as
\be
R(D^{**}_{(s)})=\frac{\Gamma\left(B \rightarrow D^{**}_{(s)} \tau \nu_\tau\right)}{\Gamma\left(B \rightarrow D^{**}_{(s)} \ell^{\prime} \nu_{\ell^{\prime}}\right)},
\en
where a large part of the theoretical and experimental uncertainties, especially the errors from the form factors, can be canceled.  One can find that most of our
predictions are comparable with other theoretical results. Compared to the S1 and S2 values given in the QCD LCSRs \cite{Gubernari:2023rfu}, our prediction for $R(D^*_0)$ gives a moderate value.

\subsection{Physical observables}
In our study of semileptonic decays, we define two additional physical observables, namely the longitudinal polarization fraction $f_L$ and the
forward-backward asymmetry $A_{FB}$, to account
for the impact of lepton mass and provide a more detailed physical picture. The results of these two physical observables are listed in Tables \ref{FL} and \ref{AFB}, respectively.
In Table {\ref{FL}}, we can clearly find that the longitudinal polarization fractions $f_L$ between
the decays $B_{(s)}\to {D}^{(\prime)}_{(s)1}e^{+}\nu_{e}$ and $B_{(s)}\to {D}^{(\prime)}_{(s)1}\mu^{+}\nu_{\mu}$ are very close to each other, 
which reflect the lepton flavor universality (LFU). In order to investigate
the dependences of the polarizations on the different $q^2$, we divide the full energy region into two segments for each decay and calculate the longitudinal polarization fractions accordingly. Region 1 is defined as $m_{\ell}^{2}<q^{2}<\frac{(m_{B_{(s)}}-m_{D^{(\prime)}_{(s1)}})^{2}+m_{\ell}^{2}}{2} $ and Region 2 is $\frac{(m_{B_{(s)}}-m_{D_{(s)1}^{(\prime)}})^{2}+m_{\ell}^{2}}{2} <q^{2}<(m_{B_{(s)}}-m_{D_{(s)1}^{(\prime)}})^{2}$.  Obviously, the longitudinal polarization fraction in Region 1 is larger than that in Region 2 for each decay. Furthermore, for the decays with $D_{(s)1}$ involved in the final states the longitudinal polarization is dominant, while it is contrary for the decays with $D^{\prime}_{(s)1}$ involved. These results can be validated by the future high-luminosity experiments.
\begin{table}[H]
	\caption{The lepton flavor universality ratios of the transitions $B_{(s)} \to D^*_0$, $D^*_{s0}$, $D^{(\prime)}_{s1}$, $D^{(\prime)}_1$.}
	\begin{center}
		\scalebox{0.9}{
			\begin{tabular}{c|c|c|c|c|c}
				\hline\hline
				Decays&Ratios&Predicted values&Decays&Ratios&Predicted values\\
				\hline
				$B^+ \to \bar{D}_{0}^{*0} \ell^+ \nu_{\ell}$&$R(\bar D^*_0)$&$0.127^{+0.003+0.003}_{-0.001-0.000}$&$B_{s}^{0}\to D_{s1}^- \ell^+ \nu_\ell$&$R(D_{s1})$&$0.102^{+0.007+0.003}_{-0.009-0.001}$\\
				\hline
				&$$&$0.063^{+0.001}_{-0.002}$ \cite{Zuo:2023ksq}&$$&$$&$0.061\pm0.003$ \cite{Zuo:2023ksq}\\
				&$$&$0.08\pm0.03$ \cite{F.U.}&$$&$$&$0.09\pm0.02$ \cite{F.U.1}\\
				\hline
				$B^0 \to D_{0}^{*-} \ell^+ \nu_{\ell}$&$R(D^*_0)$&$0.127^{+0.001+0.002}_{-0.001-0.002}$&$B_{s}^{0}\to D_{s1}^{\prime-} \ell^+ \nu_{\ell}$&$R(D^{\prime}_{s1})$&$0.111^{+0.014+0.009}_{-0.029-0.020
				}$\\
				\hline
				&$$&$0.099\pm0.015$ \cite{A.L.}&$$&$$&$0.084^{+0.001}_{-0.002}$ \cite{Zuo:2023ksq}\\
				&$$&$0.11^{+0.03}_{-0.01} \textnormal{(S1)}, 0.16^{+0.04}_{-0.02}\textnormal{(S2)}^1$ \cite{Gubernari:2023rfu}	&$$&$$&$0.07\pm0.03$ \cite{F.U.1}\\
				\hline
				$B^0_s \to D_{s0}^{*} \ell^+\nu_{\ell}$&$R(D^*_{s0})$&$0.147^{+0.003+0.001}_{-0.004-0.003}$&$B^{+}\to \bar{D}_1^{0} \ell^+ \nu_{\ell}$&$R(D_1)$&$0.091^{+0.001+0.003}_{-0.002-0.004}$\\
				\hline
				&$$&$0.080^{+0.001}_{-0.002}$ \cite{Zuo:2023ksq}&$$&$$&$0.064^{+0.003}_{-0.003}$ \cite{Zuo:2023ksq}\\
				&$$&$0.09\pm0.04$ \cite{F.U.1}&$$&$$&$0.10\pm0.02$ \cite{F.U.}\\
				&$$&$0.14^{+0.07}_{-0.02}$ \cite{Gubernari:2023rfu}&$$&$$&$0.098\pm0.007$ \cite{A.L.}\\
				\hline
				&$$&$$&$B^{+}\to \bar{D}_1^{\prime0} \ell^+ \nu_{\ell}$&$R(D^{\prime}_1)$&$0.110^{+0.030+0.051}_{-0.010-0.025}$\\
                \hline
				&$$&$$&$$&$$&$0.076^{+0.001}_{-0.002}$ \cite{Zuo:2023ksq}\\
				&$$&$$&$$&$$&$0.05\pm0.02$ \cite{F.U.}\\	
				&$$&$$&$$&$$&$0.074\pm0.012$ \cite{A.L.}\\	
				\hline\hline
			\end{tabular}\label{R}
		}
	\end{center}
	{\footnotesize $^1$ The definitions of S1 and S2 are given in Table \ref{tab2}. }
\end{table}  

\begin{table}[H]
	\caption{The longitudinal polarization fractions $f_{L}$ for the decays $B_s \to D^{(\prime)-}_{s1}\ell^{+}\nu_{\ell}$ and $B^+ \to \bar{D}^{(\prime)0}_{1}\ell^{+}\nu_{\ell}$ in Region 1 and Region 2.}
	\begin{center}
		\scalebox{0.9}{
			\begin{tabular}{|c|c|c|c||c|c|c|c|}
				\hline\hline
				Observables&Region 1&Region 2&Total&Observables&Region 1&Region 2&Total\\
				\hline\hline
				$f_{L}(B_{s}^{0}\to D_{s1}^{-} e^{+}\nu_{e})$&$0.81$&$0.50$&$0.71$&$f_{L}(B_{s}^{0}\to D_{s1}^{\prime-} e^{+}\nu_{e})$&$0.42$&$0.12$&$0.31$\\
				\hline
				$f_{L}(B_{s}^{0}\to D_{s1}^- \mu^{+}\nu_{\mu})$&$0.81$&$0.50$&$0.71$&$f_{L}(B_{s}^{0}\to D_{s1}^{\prime-} \mu^{+}\nu_{\mu})$&$0.42$&$0.12$&$0.31$\\
				\hline
				$f_{L}(B_{s}^{0}\to D_{s1}^- \tau^{+}\nu_{\tau})$&$0.66$&$0.48$&$0.56$&$f_{L}(B_{s}^{0}\to D_{s1}^{\prime-} \tau^{+}\nu_{\tau})$&$0.23$&$0.24$&$0.24$\\
				\hline\hline
				$f_{L}(B^{+}\to \bar{D}_{1}^{0} e^{+}\nu_{e})$&$0.81$&$0.50$&$0.72$&$f_{L}(B^{+}\to \bar{D}_{1}^{\prime0}e^{+}\nu_{e})$&$0.54$&$0.11$&$0.39$\\
				\hline
				$f_{L}(B^{+}\to \bar{D}_{1}^{0} \mu^{+}\nu_{\mu})$&$0.81$&$0.50$&$0.72$&$f_{L}(B^{+}\to \bar{D}_{1}^{\prime0} \mu^{+}\nu_{\mu})$&$0.54$&$0.11$&$0.39$\\
				\hline
				$f_{L}(B^{+}\to \bar{D}_{1}^{0} \tau^{+}\nu_{\tau})$&$0.67$&$0.49$&$0.56$&$f_{L}(B^{+}\to \bar{D}_{1}^{\prime0}\tau^{+}\nu_{\tau})$&$0.32$&$0.25$&$0.28$\\
				\hline\hline
			\end{tabular}\label{FL}}
	\end{center}
\end{table}

\begin{figure}[H]
	\vspace{0.32cm}
	\centering
	\subfigure[]{\includegraphics[width=0.23\textwidth]{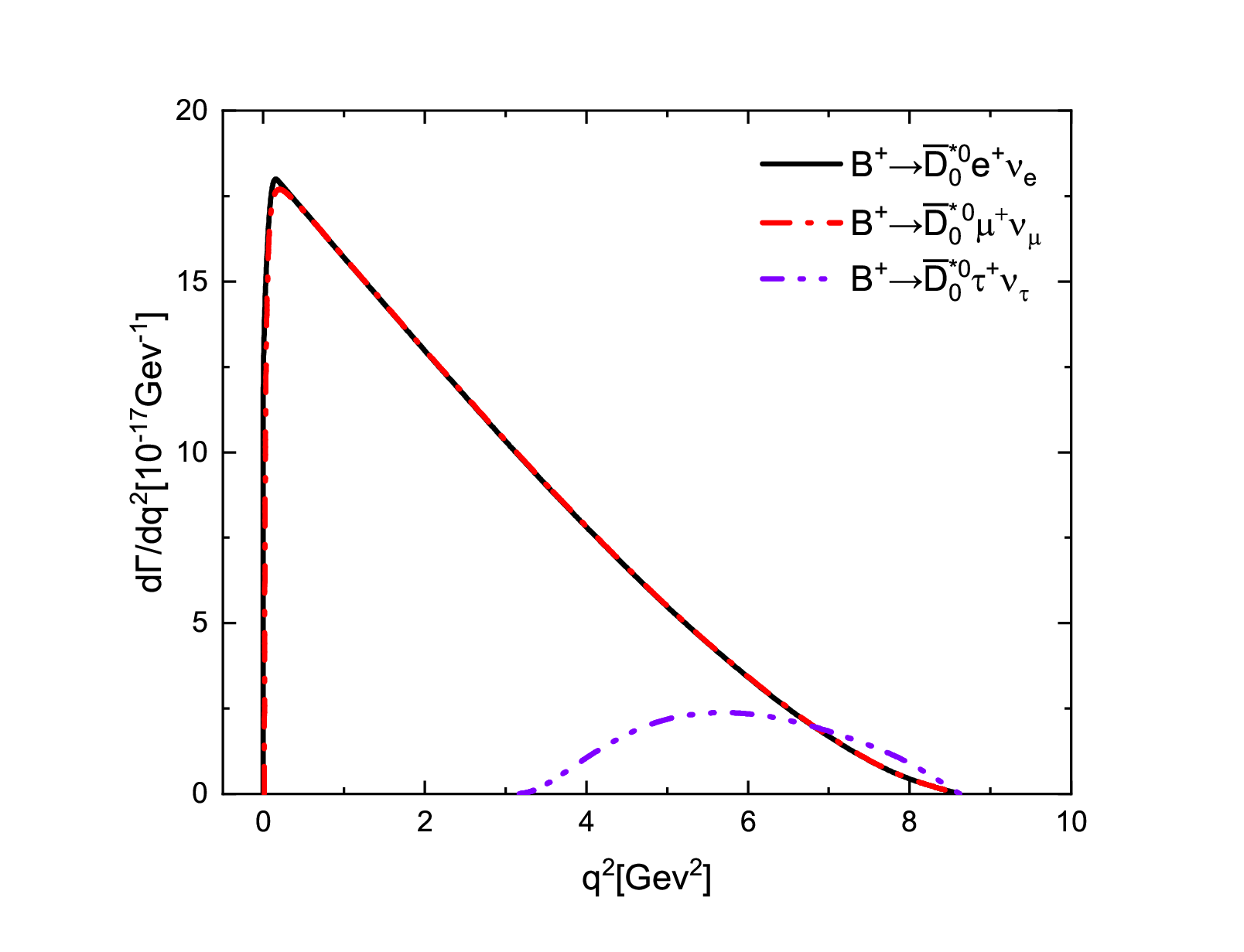}}
	\subfigure[]{\includegraphics[width=0.23\textwidth]{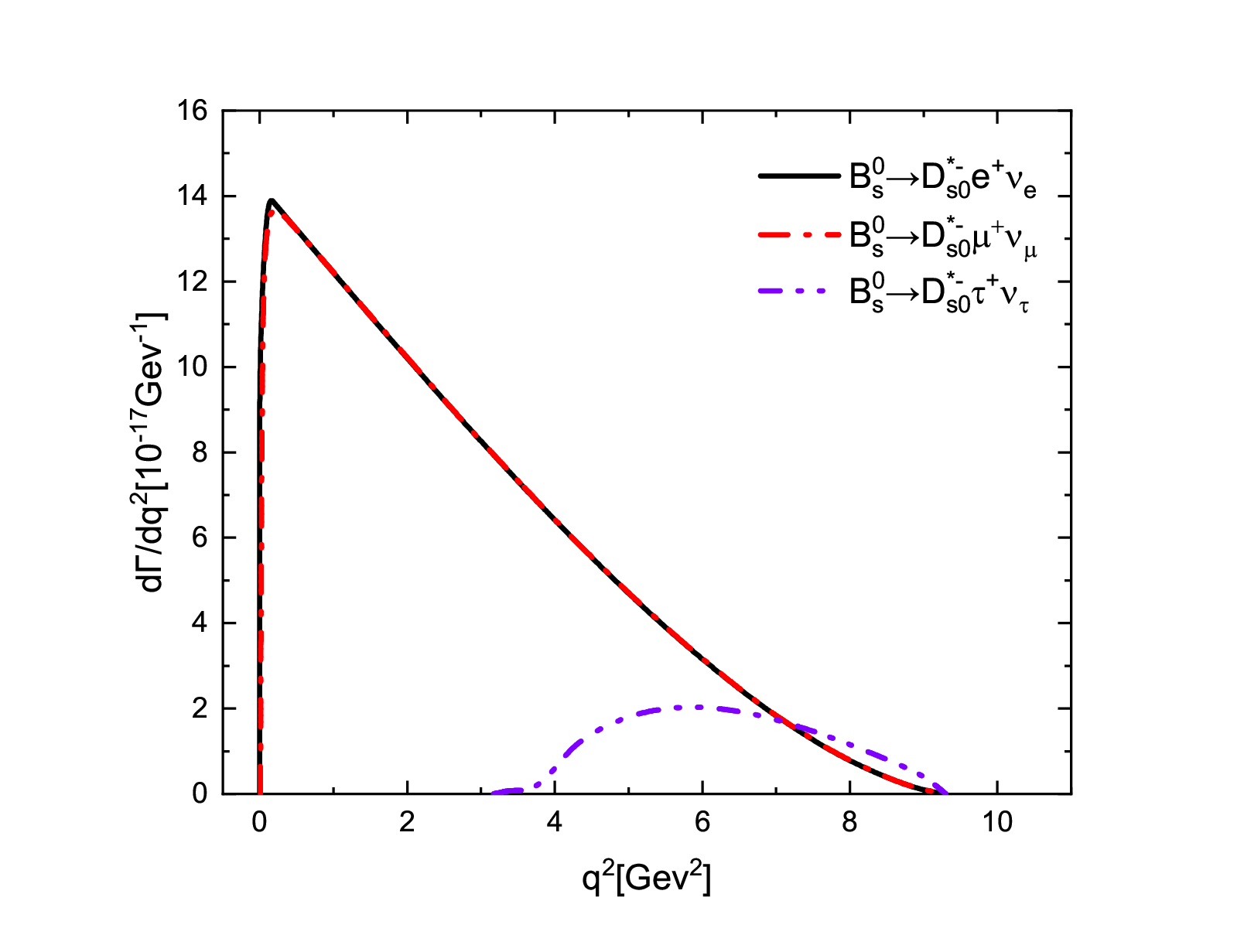}}
	\subfigure[]{\includegraphics[width=0.23\textwidth]{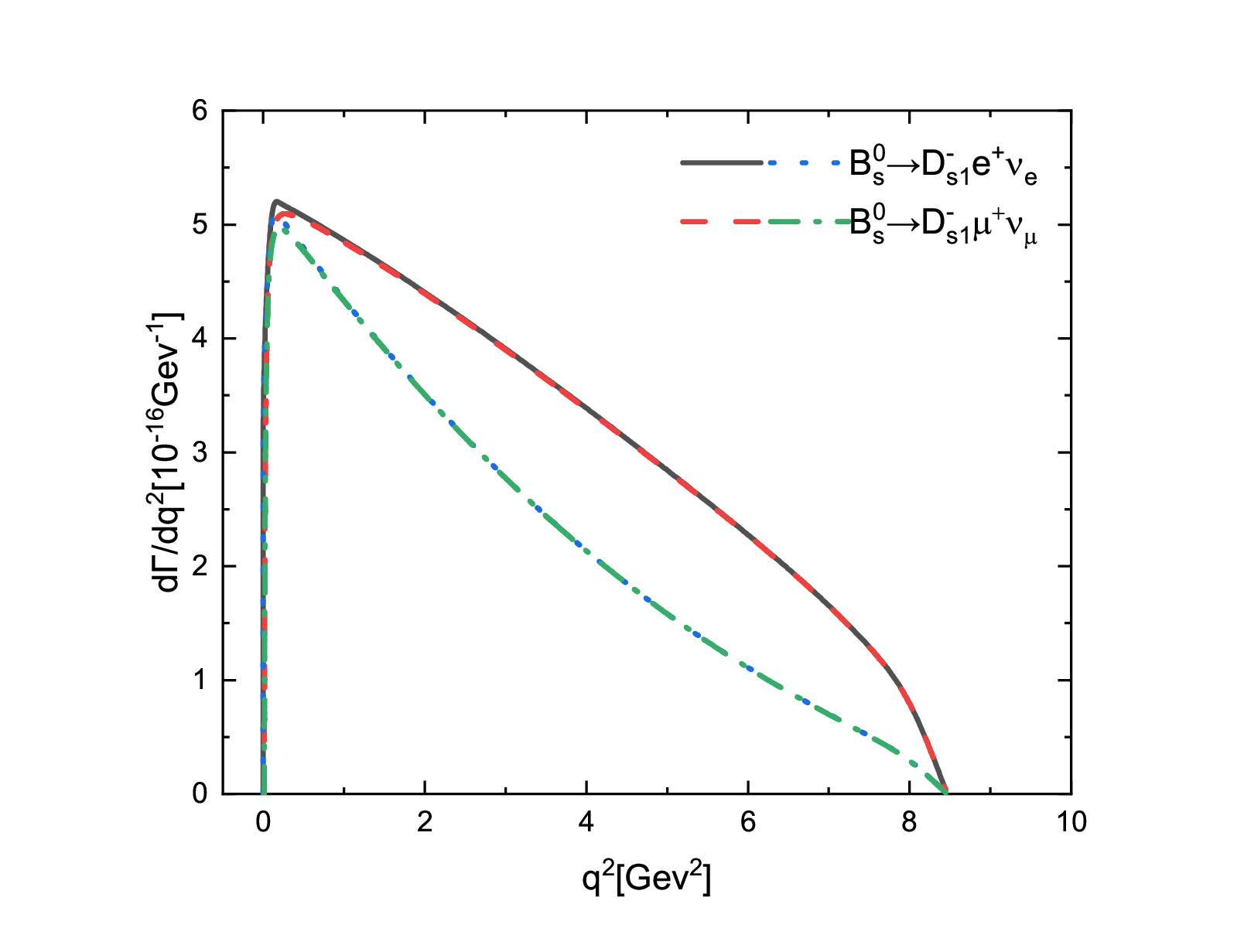}}
	\subfigure[]{\includegraphics[width=0.23\textwidth]{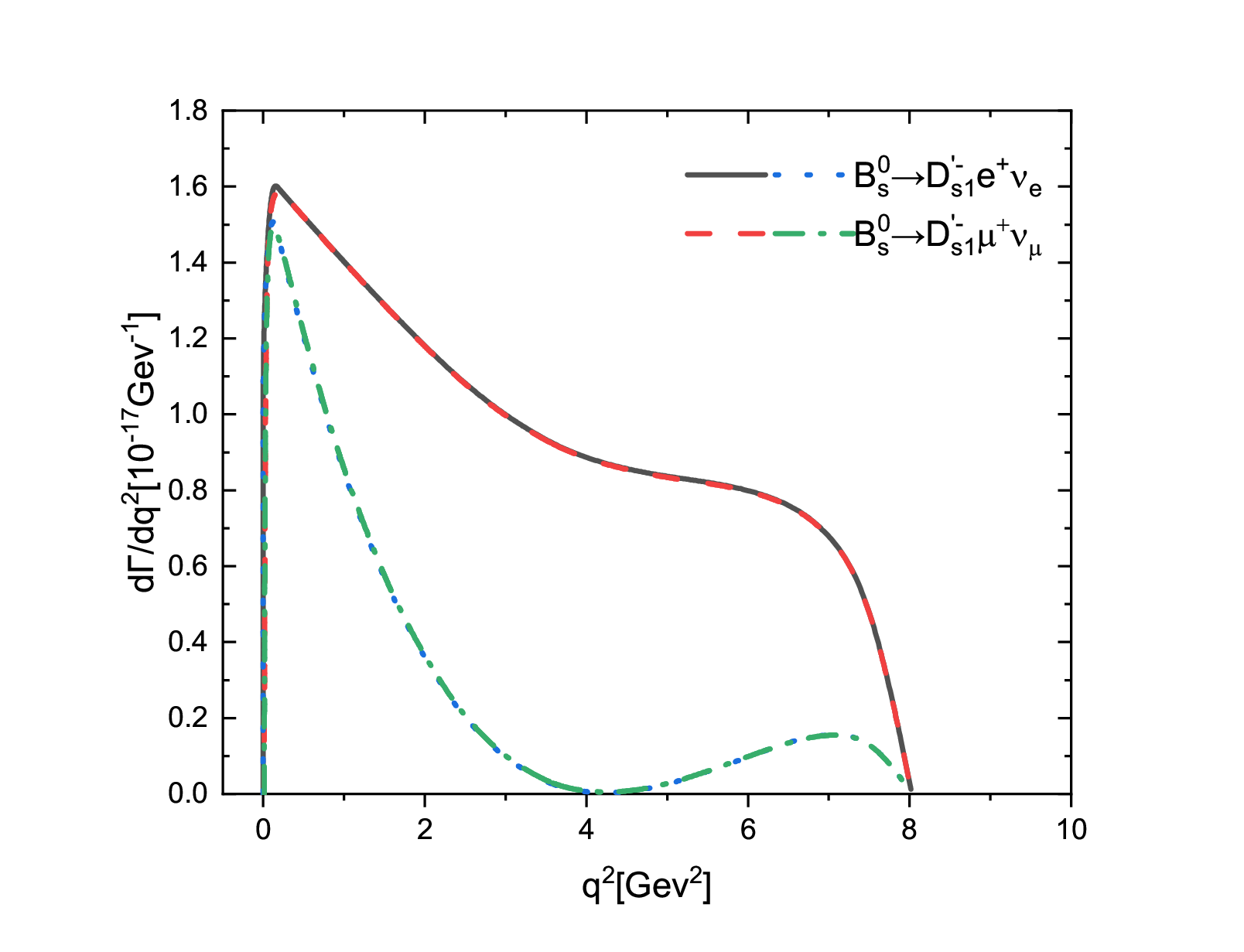}}\\
	\subfigure[]{\includegraphics[width=0.23\textwidth]{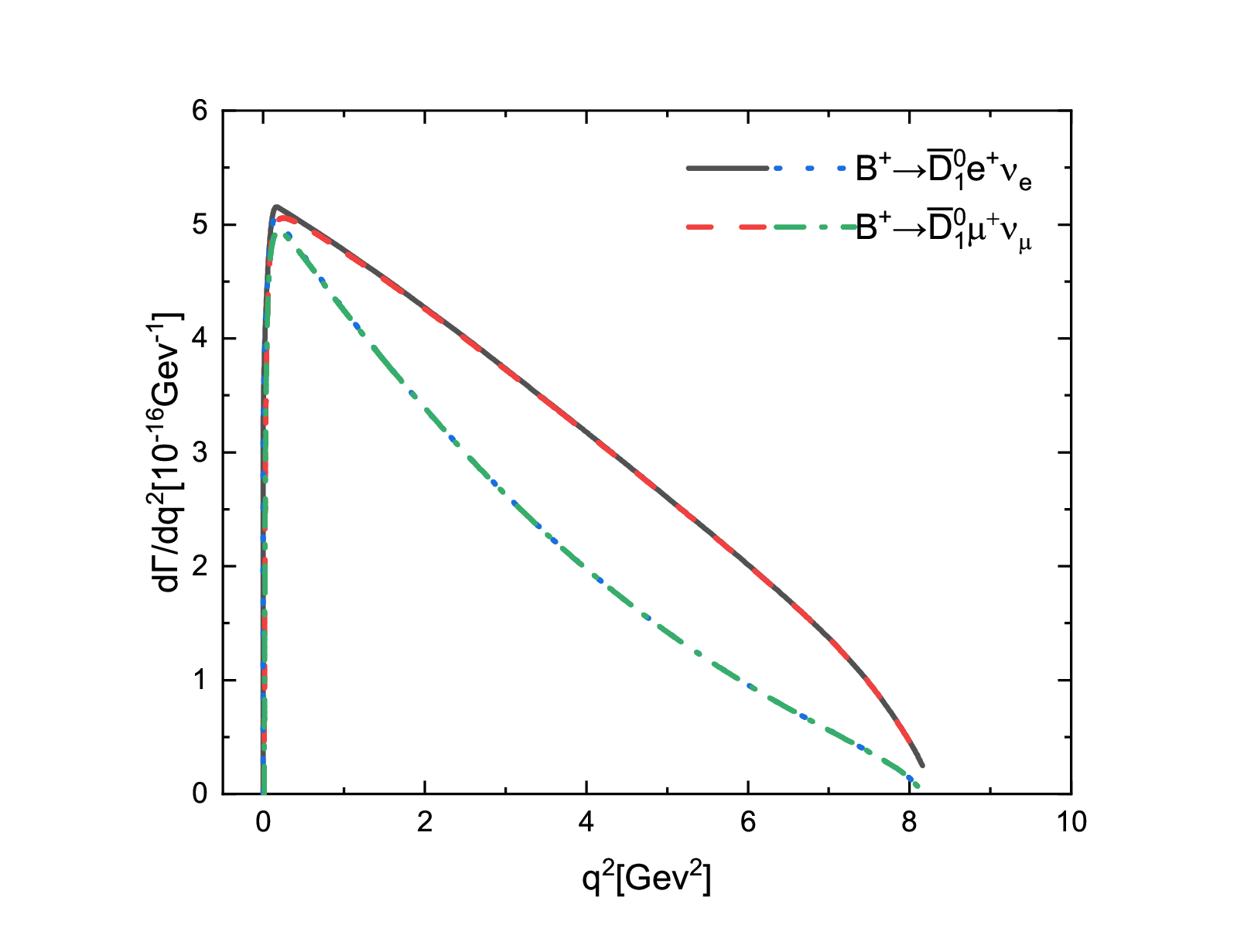}}
	\subfigure[]{\includegraphics[width=0.23\textwidth]{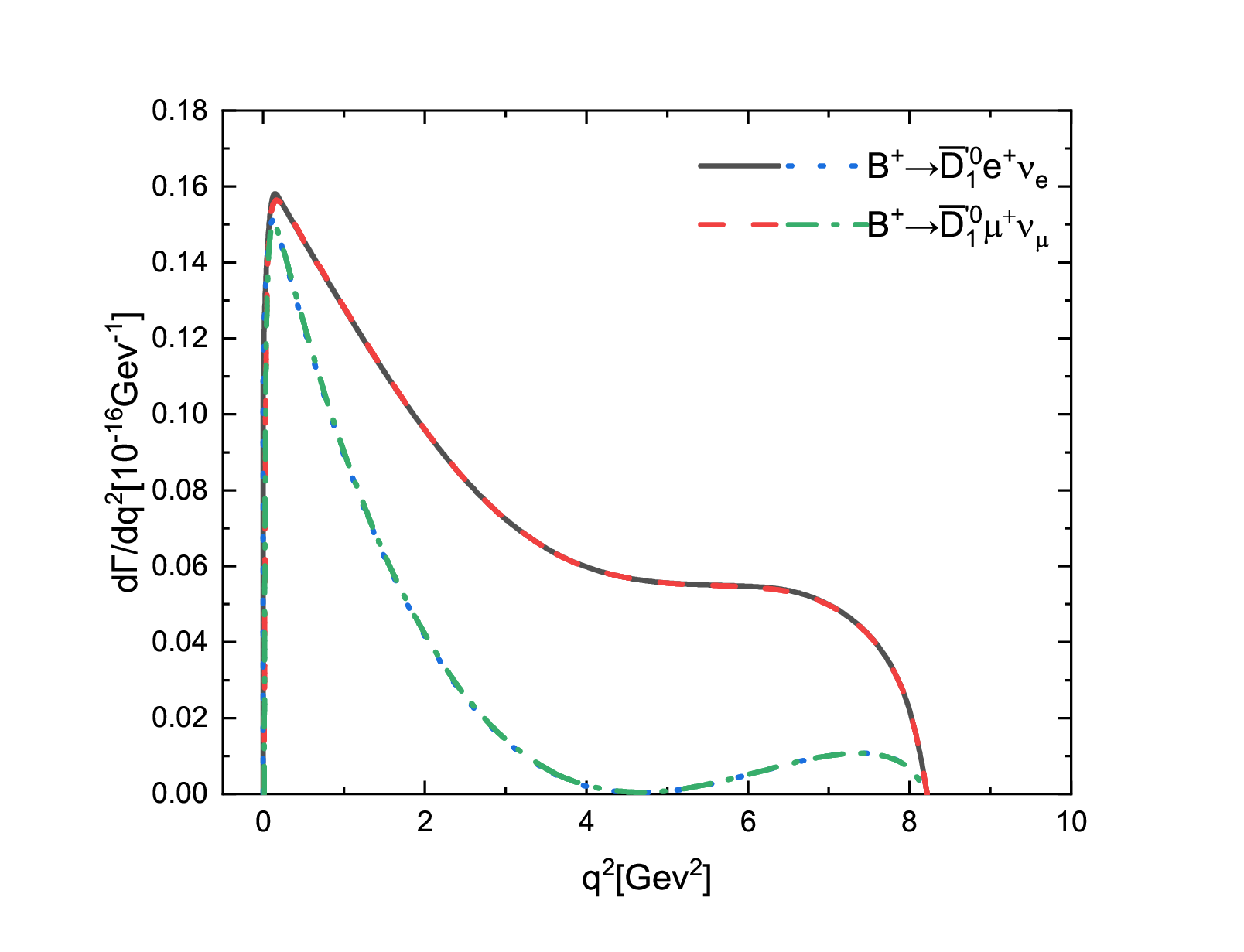}}
	\subfigure[]{\includegraphics[width=0.23\textwidth]{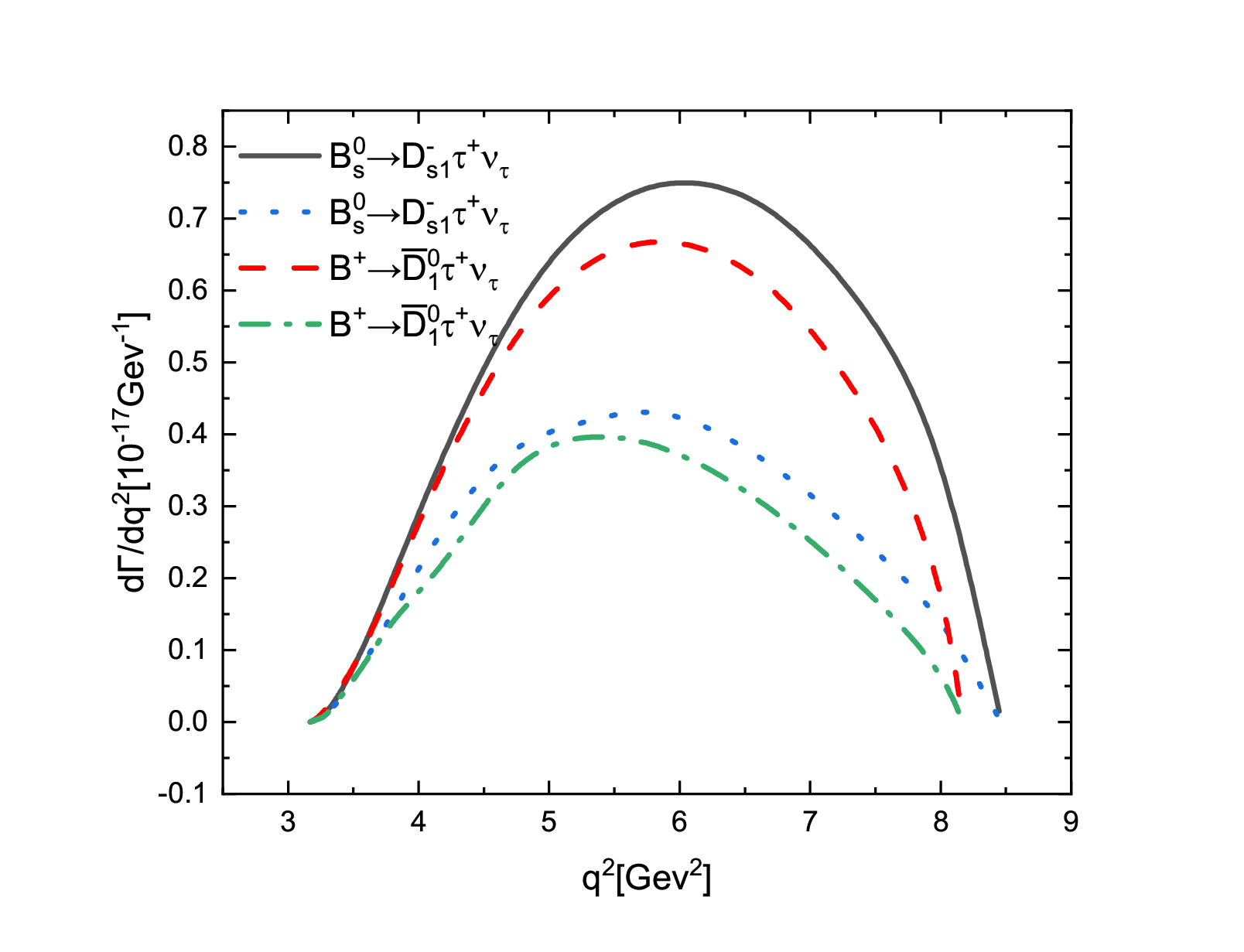}}
	\subfigure[]{\includegraphics[width=0.23\textwidth]{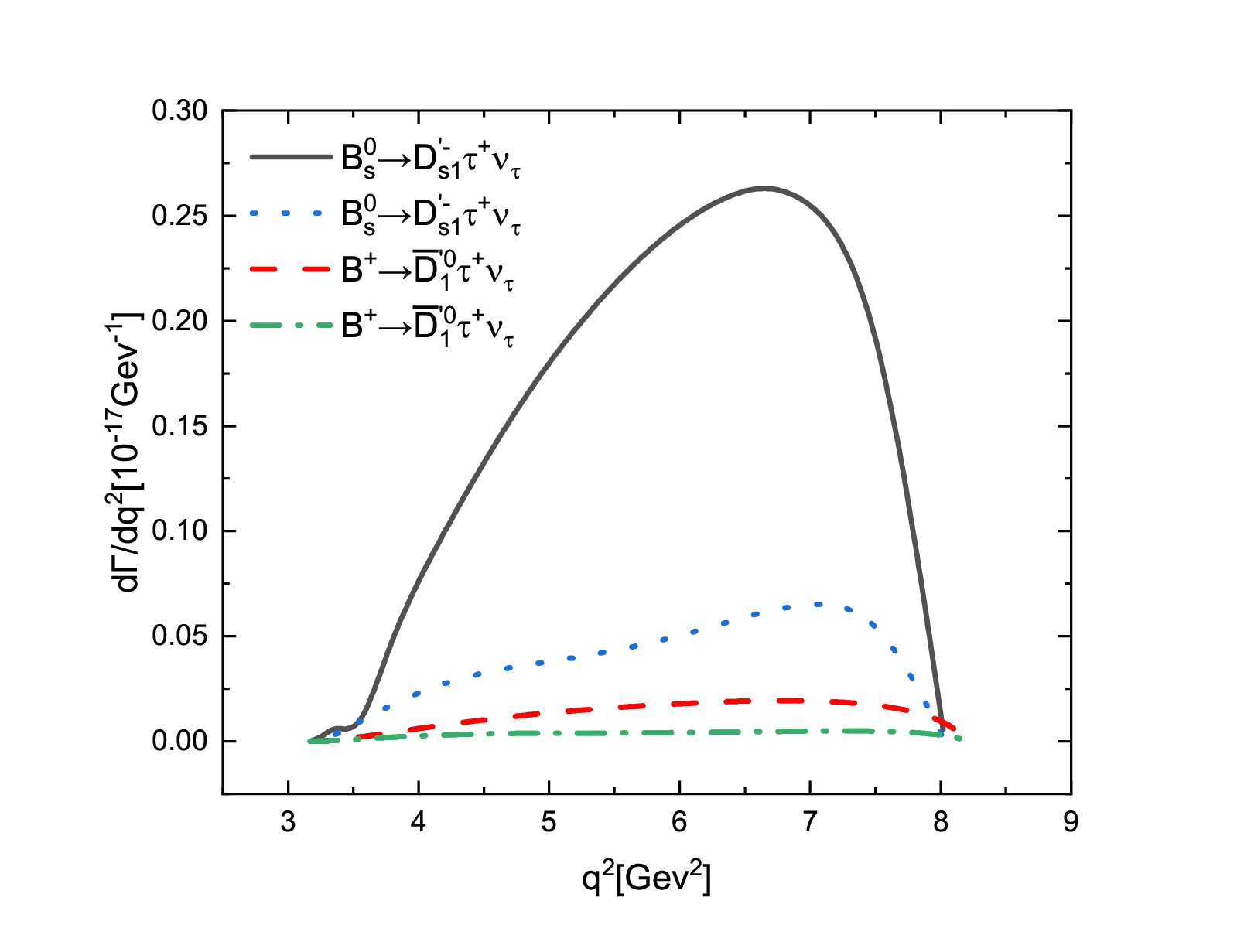}}
	\caption{ The $q^2$ dependencies of differential decay rates $d\Gamma/dq^2$ and
		$d\Gamma^L/dq^2$ for the decays $B_{(s)} \to D^*_{(s)0} \ell {\nu}_{\ell}$ and 
		$B_{(s)}\to {D}^{(\prime)}_{(s)1}\ell \nu_{\ell}$.}\label{fig:dq}
\end{figure}

In Figure \ref{fig:dq}, we plot the $q^2$ dependencies of the differential decay rates for the channels $B_{(s)} \to D^*_{(s)0} \ell {\nu}_{\ell}$ and 
$B_{(s)}\to {D}^{(\prime)}_{(s)1}\ell \nu_{\ell}$. One can find that the line shapes of these differential distributions are constrained by the phase space, in other words, the lepton mass. The polarization needs to be considered in the decays $B_{s}\to {D}^{(\prime)}_{s1}\ell \nu_{\ell}$ and $B\to D^{(\prime)}_{1}\ell \nu_{\ell}$, which is shown in Figures \ref{fig:dq}(c)-\ref{fig:dq}(h). It is obvious that for the decays with $D_{(s)1}$ involved the longitudinal polarization is dominant in small $q^2$ region and comparable with the transverse ones in large $q^2$ region. While for the decays with $D^\prime_{(s)1}$ involved the transverse polarizations are dominant, especially in large $q^2$ region. 
It is interesting that taking some special $q^2$ values for the decays $B\to  {D}^{\prime}_{1}\ell^{\prime} \nu_{\ell^{\prime}}$ and $B_{s}\to {D}^{\prime}_{s1}\ell^{\prime} \nu_{\ell^{\prime}}$, we can find that the contribution from the longitudinal polarization almost disappears with only the transverse polarizations left, which is shown in Figures \ref{fig:dq}(d) and \ref{fig:dq}(f). Maybe such a phenomenon can be checked in the future LHC and Super KEKB experiments to test the present mixing mechanism.

\begin{table}[H]
	\caption{The forward-backward asymmetries $A_{FB}$ for the decays $B_{(s)} \to D^*_{(s)0} \ell {\nu}_{\ell}$ and 
		$B_{(s)}\to {D}^{(\prime)}_{(s)1}\ell \nu_{\ell}$.}
	\begin{center}
		\scalebox{0.8}{
			\begin{tabular}{|c|c|c|c|c|}
				\hline\hline
				Channels & &$B^{+}\to \bar{D}^*_0e^{+}\nu_{e}$&$B^{+}\to \bar{D}^*_0\mu^{+}\nu_{\mu}$&$B^{+}\to \bar{D}^*_0 \tau^{+}\nu_{\tau}$\\
				\hline
				$A_{FB}$&This work&$(5.820^{+1.494+2.582}_{-0.904-2.108})\times10^{-7}$&$0.020^{+0.005+0.009}_{-0.003-0.007}$&$0.383^{+0.102+0.177}_{-0.063-0.143}$\\
				\hline
				Channels & &$B_{s}^{0}\to D^{*-}_{s0} e^{+}\nu_{e}$&$B_{s}^{0}\to D^{*-}_{s0} \mu^{+}\nu_{\mu}$&$B_{s}^{0}\to D^{*-}_{s0} \tau^{+}\nu_{\tau}$\\
				\hline
				\multirow{2}{*}{$A_{FB}$}&This work&$(5.465^{+1.100+2.298}_{-0.515-1.898})\times10^{-7}$&$0.019^{+0.004+0.010}_{-0.002-0.007}$&$0.381^{+0.079+0.168}_{-0.038-0.138}$\\
                $$&\cite{Albertus:2014bfa}&$8.22\times10^{-7}$&$0.016$&$0.39$\\
				\hline\hline
				Channel& &$B_{s}^{0}\to D_{s1}^- e^{+}\nu_{e}$&$B_{s}^{0}\to D_{s1}^- \mu^{+}\nu_{\mu}$&$B_{s}^{0}\to D_{s1}^- \tau^{+}\nu_{\tau}$\\
				\hline
				\multirow{2}{*}{$A_{FB}$}&This work&$-0.175^{+0.016+0.027+0.011}_{-0.014-0.016-0.017}$&$-0.174^{+0.016+0.026+0.010}_{-0.014-0.017-0.017}$&$-0.129^{+0.010+0.021+0.003}_{-0.015-0.017-0.016}$\\
                 $$&\cite{Albertus:2014bfa}&$-0.19$&$-0.18$&$0.10$\\
				\hline
				Channels& &$B_{s}^{0}\to D_{s1}^{\prime-} e^{+}\nu_{e}$&$B_{s}^{0}\to D_{s1}^{\prime-} \mu^{+}\nu_{\mu}$&$B_{s}^{0}\to D_{s1}^{\prime-} \tau^{+}\nu_{\tau}$\\
				\hline
				\multirow{2}{*}{$A_{FB}$}&This work&$-0.407^{+0.104+0.129+0.020}_{-0.101-0.185-0.029}$&$-0.405^{+0.104+0.129+0.020}_{-0.102-0.186-0.029}$&$-0.248^{+0.026+0.080+0.048}_{-0.064-0.027-0.007}$\\
                 $$&\cite{Albertus:2014bfa}&$-0.41$&$-0.40$&$-0.20$\\
				\hline\hline
				Channels& &$B^{+}\to \bar{D}_{1}^{0} e^{+}\nu_{e}$&$B^{+}\to \bar{D}_{1}^{0} \mu^{+}\nu_{\mu}$&$B^{+}\to \bar{D}_{1}^{0} \tau^{+}\nu_{\tau}$\\
				\hline
				$A_{FB}$&This work&$-0.178^{+0.014+0.018+0.008}_{-0.030-0.025-0.017}$&$-0.177^{+0.012+0.016+0.006}_{-0.032-0.028-0.019}$&$-0.131^{+0.024+0.037+0.014}_{-0.034-0.036-0.025}$\\
				\hline
				Channels& &$B^{+}\to \bar{D}_{1}^{\prime0} e^{+}\nu_{e}$&$B^{+}\to \bar{D}_{1}^{\prime0} \mu^{+}\nu_{\mu}$&$B^{+}\to \bar{D}_{1}^{\prime0} \tau^{+}\nu_{\tau}$\\
				\hline
				$A_{FB}$&This work&$-0.316^{+0.115+0.085+0.083}_{-0.190-0.090-0.048}$&$-0.314^{+0.111+0.081+0.079}_{-0.096-0.095-0.043}$&$-0.186^{-0.013+0.041+0.024}_{-0.032-0.013-0.017}$\\
				\hline\hline
			\end{tabular}\label{AFB}}
	\end{center}
\end{table}
\begin{figure}[H]
	\vspace{0.32cm}
	\centering
	\subfigure[]{\includegraphics[width=0.30\textwidth]{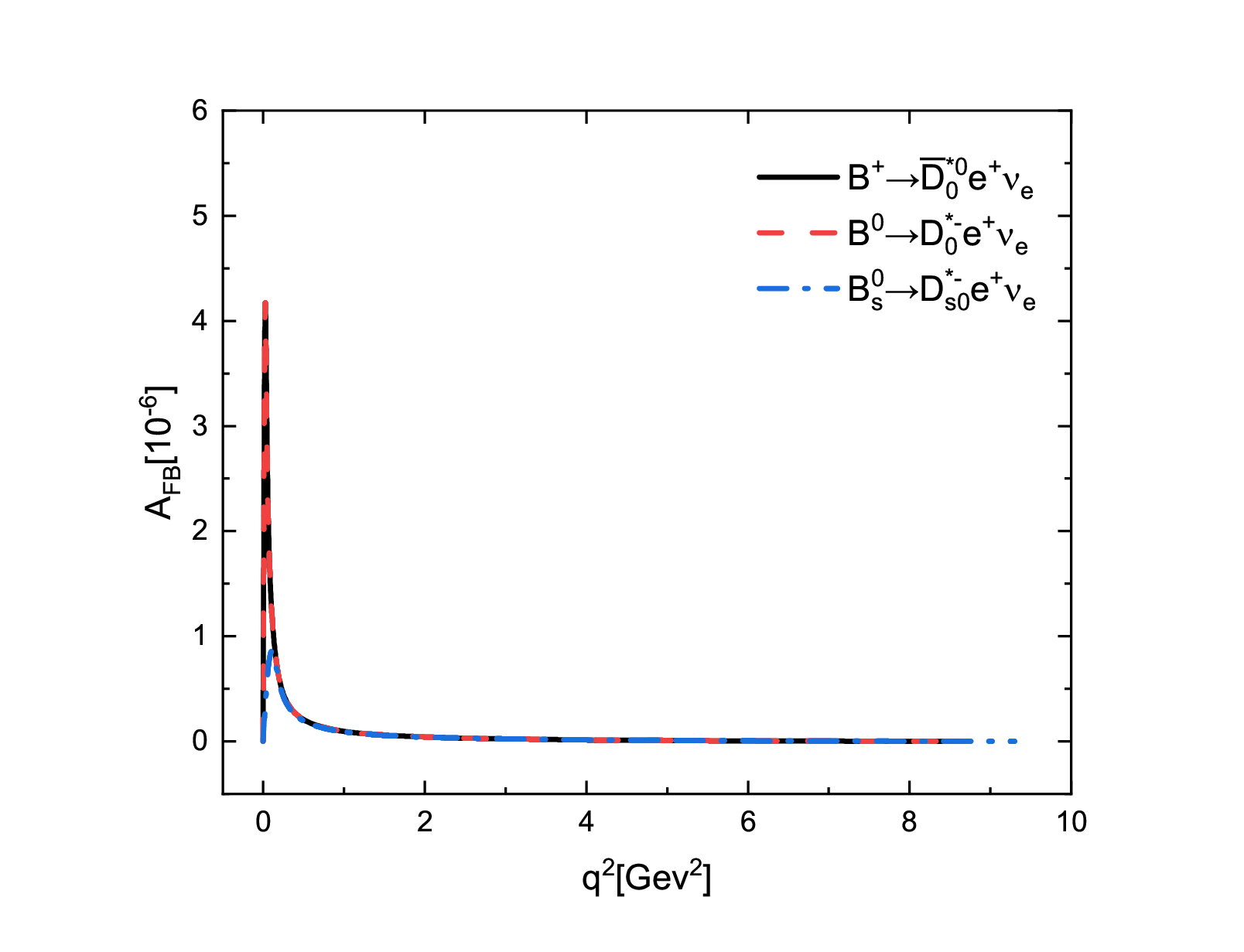}}
	\subfigure[]{\includegraphics[width=0.30\textwidth]{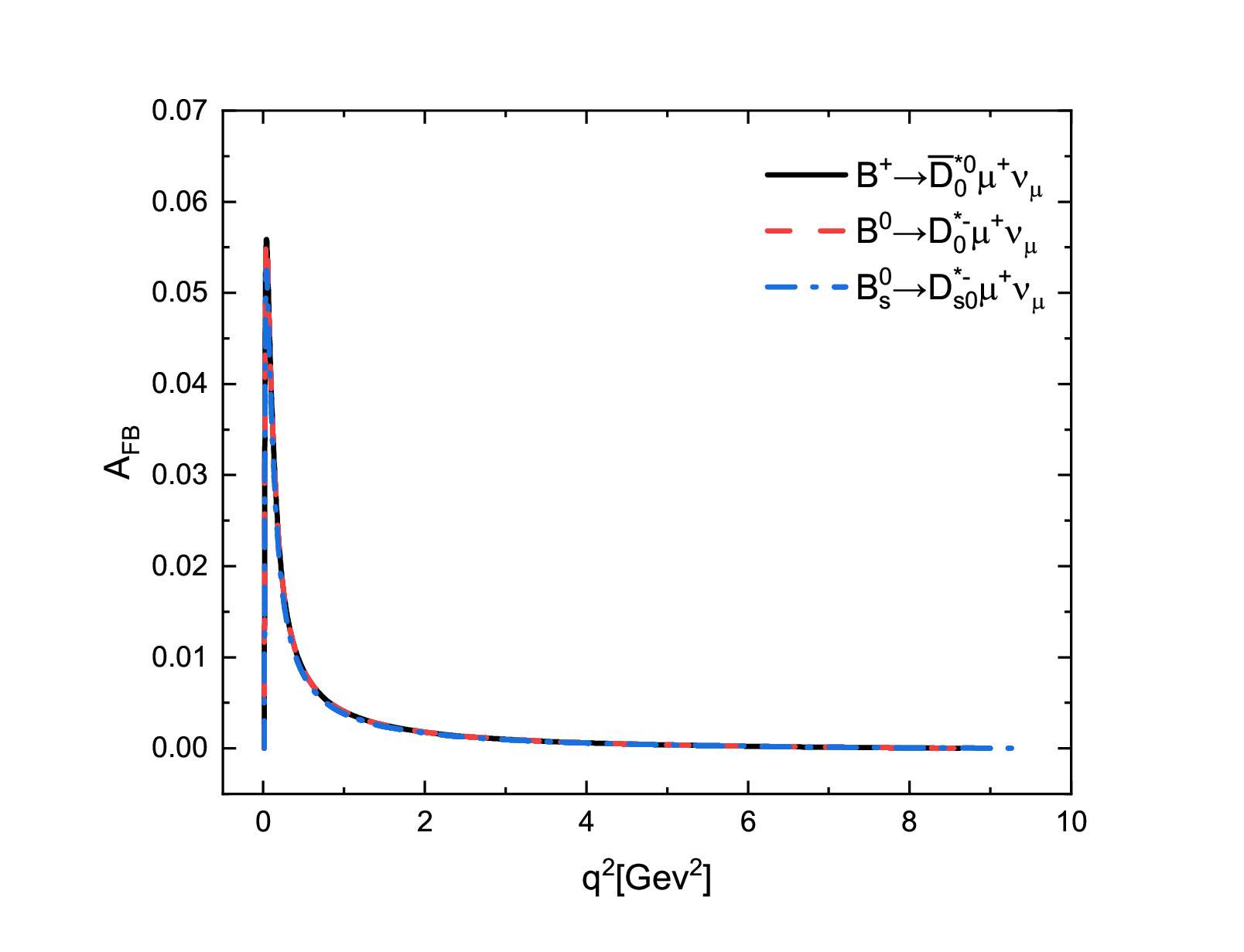}}
	\subfigure[]{\includegraphics[width=0.30\textwidth]{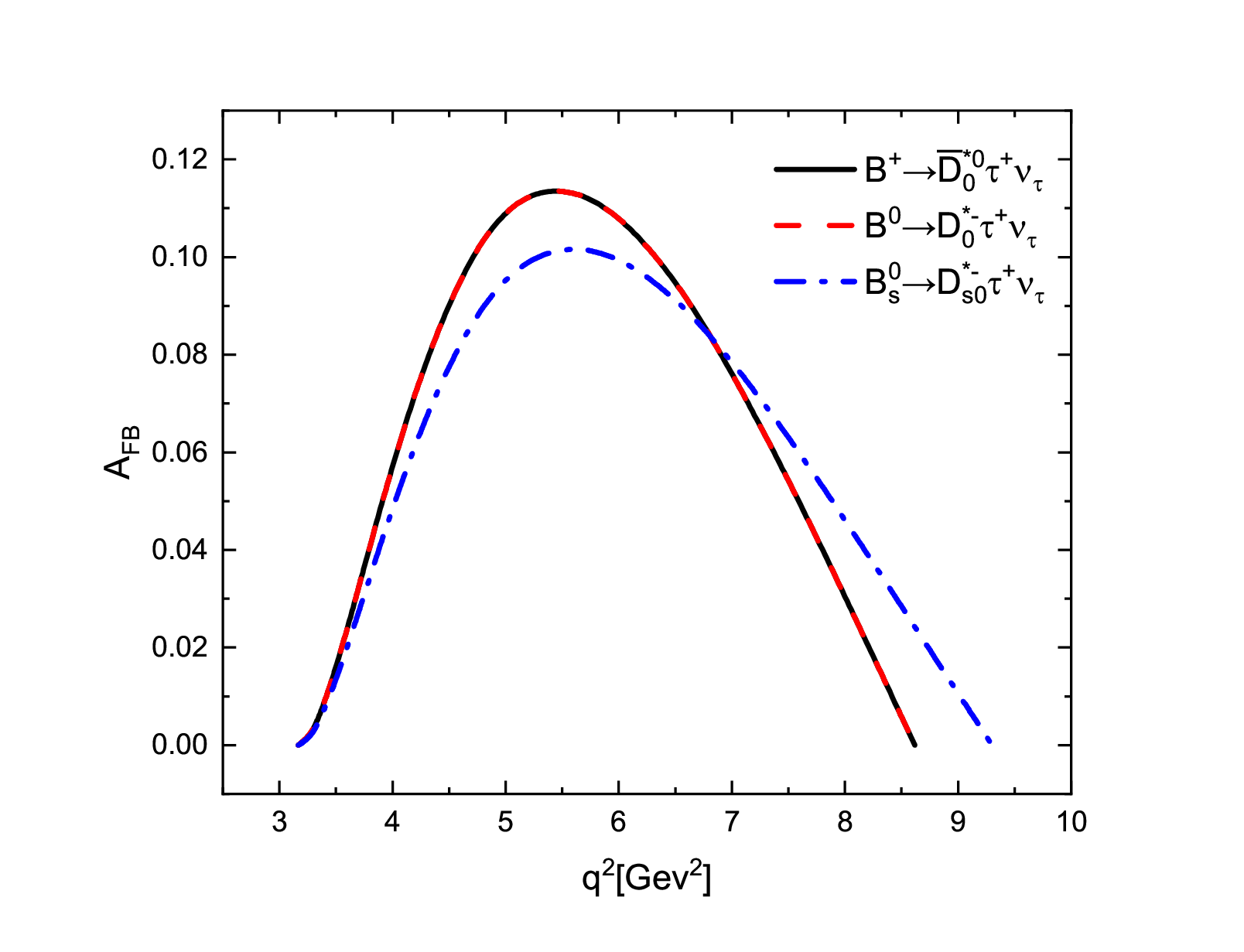}}\\
	\subfigure[]{\includegraphics[width=0.30\textwidth]{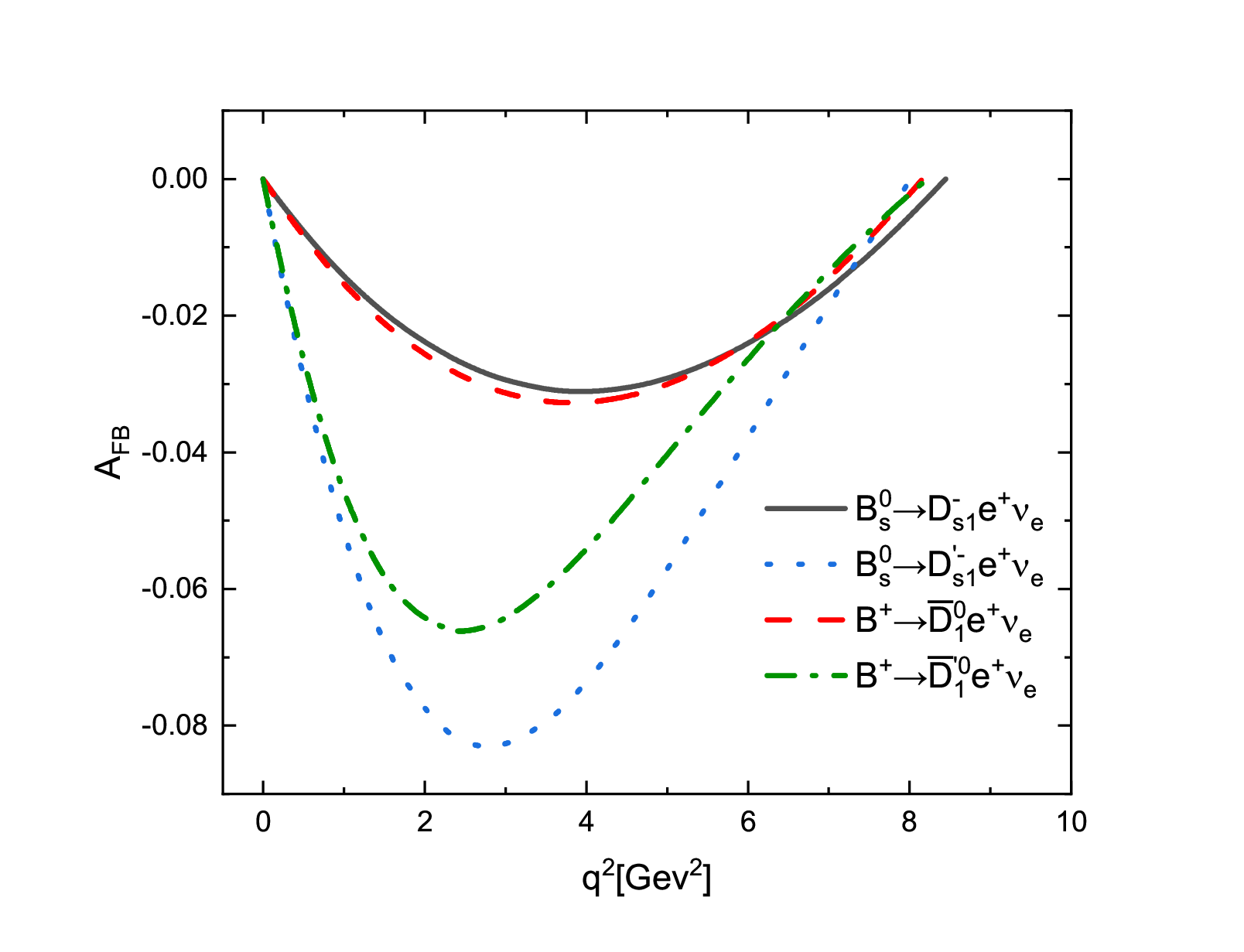}}
	\subfigure[]{\includegraphics[width=0.30\textwidth]{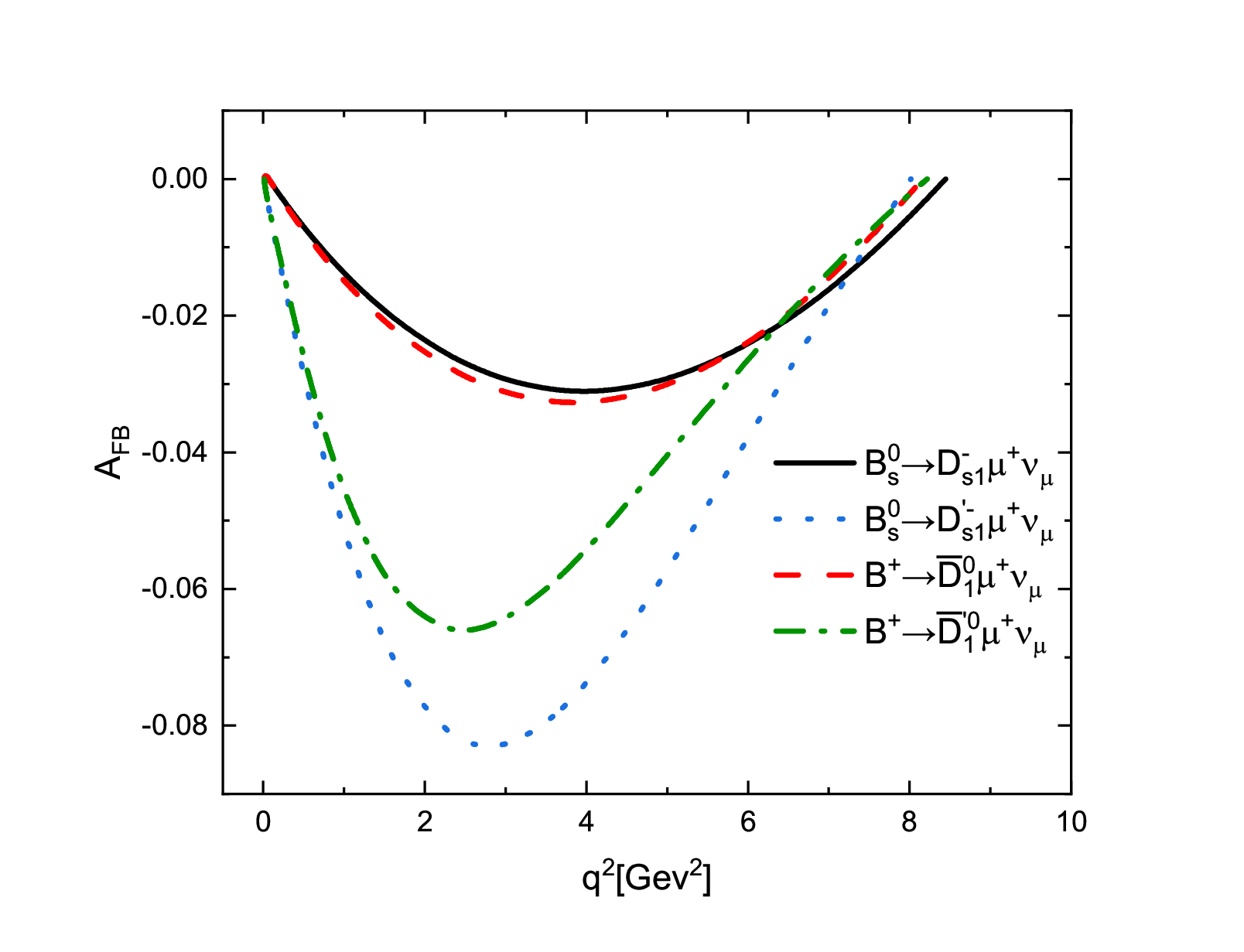}}
	\subfigure[]{\includegraphics[width=0.30\textwidth]{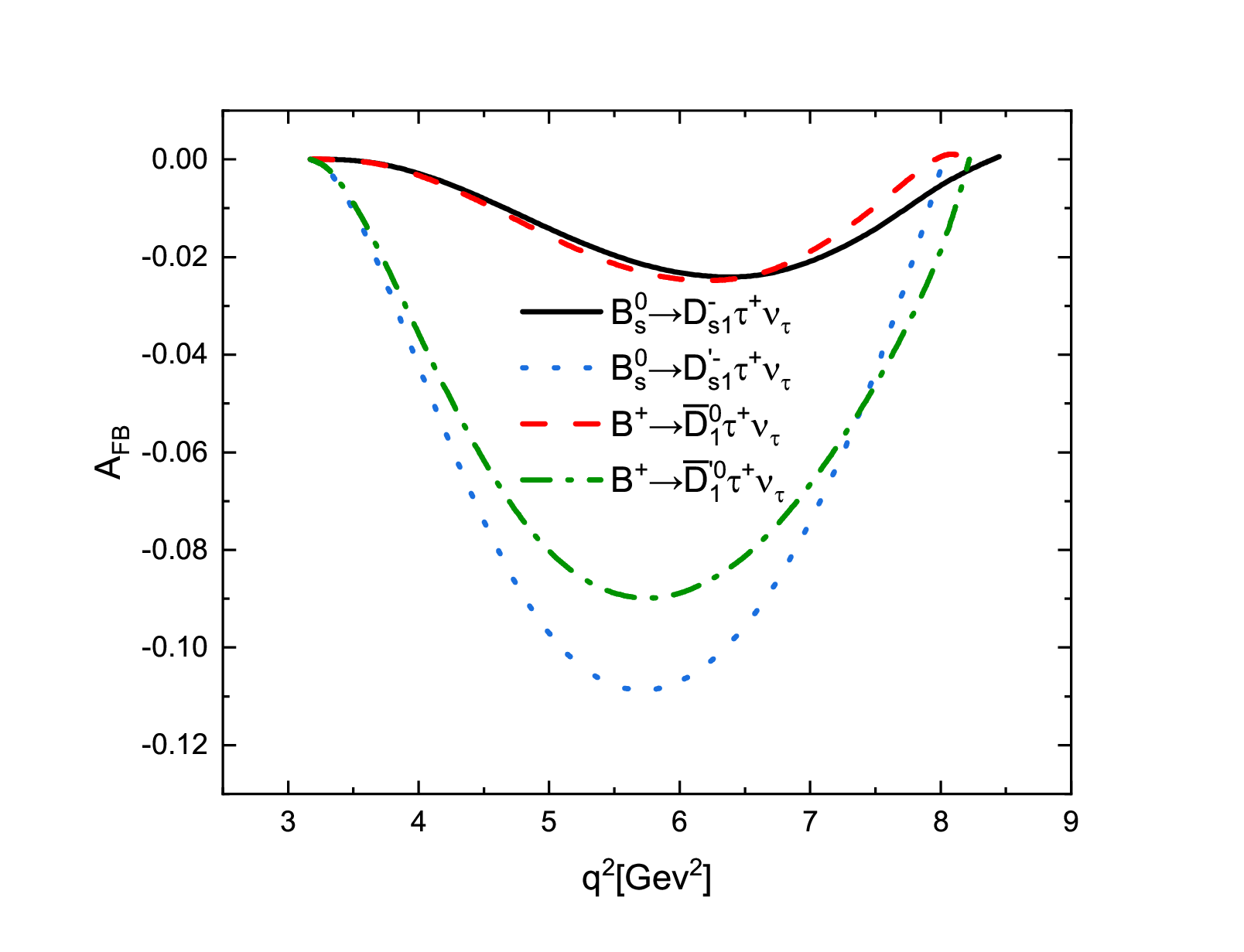}}
	\caption{The $q^2$ dependencies of the forward-backward asymmetries
		$A_{FB}$ for the decays $B_{(s)} \to D^*_{(s)0} \ell {\nu}_{\ell}$ and 
		$B_{(s)}\to {D}^{(\prime)}_{(s)1}\ell \nu_{\ell}$.}\label{fig:AFB}
\end{figure}

From Table \ref{AFB}, we find that the ratios of the forward-backward
asymmetries $A^{\mu}_{FB}$/ $A^{e}_{FB}$ between the semileptonic decays $B_{(s)}\to {D}^*_{(s)0}\mu^{+}\nu_{\mu}$ and $B_{(s)} \to {D}^*_{(s)0}e^{+}\nu_{e}$ 
are about $3.5\times10^{4}$.  The reason
is that the forward-backward asymmetries $A_{FB}$ for the decays  $B_{(s)}\to {D}^*_{(s)0}\ell^{+}\nu_{\ell}$ are proportional to the square of the lepton mass. Undoubtedly, the effect of lepton mass
can be well checked in such decay mode with a scalar meson involved in the final states. While for the decays
$B_{(s)}\to {D}^{(\prime)}_{(s)1}\ell^\prime\nu_{\ell^\prime}$, the values of the forward-backward asymmetries $A^{\mu}_{FB}$ and $A^{e}_{FB}$ are
almost equal to each other. The magnitudes of the $A_{FB}$ for the decays $B_{(s)}\to {D}^{\prime}_{(s)1}\ell\nu_{\ell}$ are larger than those for the decays $B_{(s)}\to {D}_{(s)1}\ell\nu_{\ell}$. It is worth mentioning that our results are consistent well with those calculated in  the nonrelativistic constituent quark models \cite{Albertus:2014bfa}, which are shown in Table \ref{AFB}.
In Figure \ref{fig:AFB}, we also display the $q^2$-dependencies of the forward-backward asymmetries $A_{FB}$ for the decays $B_{(s)} \to D^*_{(s)0} \ell {\nu}_{\ell}$ and 
$B_{(s)}\to {D}^{(\prime)}_{(s)1}\ell \nu_{\ell}$. Obviously, the signs of the $A_{FB}$ for the decays $B_{(s)} \to D^*_{(s)0} \ell {\nu}_{\ell}$ are contrary with those for the decays $B_{(s)}\to {D}^{(\prime)}_{(s)1}\ell \nu_{\ell}$. The lepton mass effects can be easily  observed in these figures.
\section{Summary}\label{sum}
In this work, we used the covariant light-front quark method to comprehensively investigate the
semileptonic  $B_{(s)}$ decays to $D_{0}^{\ast}$, $D_{s0}^{\ast}$, $D^{(\prime)}_{s1}$ and $D^{(\prime)}_{1}$, which can provide an important reference for future experiments.  
We calculated the branching ratios, the longitudinal polarization fractions $f_L$, and the forward-backward asymmetries $A_{FB}$ for these semileptonic $B_{(s)}$ decays using the helicity amplitudes combined with form factors. We found the following points:
\begin{enumerate}
\item The small form factors of the transitions $B_{(s)} \to D_{0}^{\ast}, D_{s0}^{\ast}$ are related to the small decay constants $f_{D_{0}^{\ast}}$ and $f_{D_{s0}^{\ast}}$. Unfortunately, there are large uncertainties in these two
decay constants. Combined with the data, our predictions for the branching ratios of the semileptonic $B_{(s)}$ meson decays with $D_{0}^{\ast}$ and $ D_{s0}^{\ast}$ involved are helpful in probing the inner structures of these two resonances. Recently, Belle updated their measurement for the decays $B^{0}\to D^{*-}_0 \ell^{\prime+}\nu_{\ell^\prime}$  with only a small upper limit $Br(B^{0}\to D^{*-}_0 \ell^{\prime+}\nu_{\ell^\prime})<0.44\times10^{-3}$ obtained, which is much larger than most theoretical predictions. We urge our experimental colleagues to perform further more precise measurements to clarify this new puzzle.

\item In our considered decays, the branching ratios of the channels $B^0_s\to D^-_{s1}\ell^+ {\nu}_{\ell}$ are (much) lager than those of the decays $B^0_s\to D^{\prime-}_{s1}\ell^+ \nu_{\ell}$. This is because the related form factors of the transition $B_s\to D_{s1}$ are much larger than those of the transition $B_s\to D^{\prime}_{s1}$. There exists a similar situation between the decays $B^+\to \bar{D}^{0}_{1}\ell^+ \nu_{\ell}$ and $B^+\to \bar{D}^{\prime0}_{1}\ell^+ \nu_{\ell}$. In addition, we calculated the dependencies of the branching ratios of the decays $B\to D^{(\prime)}_1 \ell\nu_{\ell}$ and $B_s\to  D^{(\prime)}_{s1} \ell\nu_{\ell}$ on the mixing angle $\theta_s$. One can find that taking some negative mixing angle $\theta_s$ values within a range from $-30.3^\circ$ to $-24.9^\circ$ can explain the data, which correspond to $\theta$ within the range $5^\circ\sim10.4^\circ$.

\item  In these semileptonic decays $B_{(s)}\to D^{(\prime)}_{(s)1}\ell \nu_{\ell}$, the longitudinal polarization fractions in small $q^2$ region are always larger than those in large $q^2$ region. Furthermore, the longitudinal polarization for the decays with $D_{(s)1}$ involved in the final states is dominant, while it is contrary for the decays with $D^{\prime}_{(s)1}$ involved. It is interesting that taking some special $q^2$ values for the decays $B\to D^{\prime}_{1}\ell^{\prime} \nu_{\ell^{\prime}}$ and $B_{s}\to {D}^{\prime}_{s1}\ell^{\prime} \nu_{\ell^{\prime}}$, we find that the contribution from the longitudinal polarization almost disappears with only the transverse polarizations left. Maybe such a phenomenon can be searched for in the future LHC and Super KEKB experiments to test the present mixing mechanism. 
\end{enumerate}
\section*{Acknowledgment}
We thank Prof. Guo-Li Wang for helpful discussions. This work is partly supported by the National Natural Science
Foundation of China under Grant No. 11347030, the Program of
Science and Technology Innovation Talents in Universities of Henan
Province 14HASTIT037, as well as the Natural Science Foundation of Henan
Province under Grant No. 232300420116.
\appendix
\section{Some specific rules under the $p^-$ intergration}
When preforming the integraion, we need to include the zero-mode contributions. It amounts to performing the integration in a proper way in the CLFQM. Specificlly we
use the following rules given in Refs. \cite{Cheng:2003sm,Jaus}
\be \hat{p}_{1 \mu}^{\prime} &\doteq &   P_{\mu}
A_{1}^{(1)}+q_{\mu} A_{2}^{(1)},\\
\hat{p}_{1 \mu}^{\prime}
\hat{p}_{1 \nu}^{\prime}  &\doteq & g_{\mu \nu} A_{1}^{(2)} +P_{\mu}
P_{\nu} A_{2}^{(2)}+\left(P_{\mu} q_{\nu}+q_{\mu} P_{\nu}\right)
A_{3}^{(2)}+q_{\mu} q_{\nu} A_{4}^{(2)},\\
Z_{2}&=&\hat{N}_{1}^{\prime}+m_{1}^{\prime 2}-m_{2}^{2}+\left(1-2
x_{1}\right) M^{\prime 2} +\left(q^{2}+q \cdot P\right)
\frac{p_{\perp}^{\prime} \cdot q_{\perp}}{q^{2}},\\
A_{1}^{(1)}&=&\frac{x_{1}}{2}, \quad A_{2}^{(1)}=
A_{1}^{(1)}-\frac{p_{\perp}^{\prime} \cdot q_{\perp}}{q^{2}},\quad A_{3}^{(2)}=A_{1}^{(1)} A_{2}^{(1)},\\
A_{4}^{(2)}&=&\left(A_{2}^{(1)}\right)^{2}-\frac{1}{q^{2}}A_{1}^{(2)},\quad A_{1}^{(2)}=-p_{\perp}^{\prime 2}-\frac{\left(p_{\perp}^{\prime}
\cdot q_{\perp}\right)^{2}}{q^{2}}, \quad A_{2}^{(2)}=\left(A_{1}^{(1)}\right)^{2}.  \en
\section{EXPRESSIONS OF $B \rightarrow D^*_0, \;^iD_{1}$ FORM FACTORS}
\begin{footnotesize}
	\begin{eqnarray}
		F_{1}^{B D^*_0}\left(q^{2}\right)  &=& \frac{N_{c}}{16 \pi^{3}} \int d x_{2} d^{2} p_{\perp}^{\prime} \frac{h_{B}^{\prime}
			h_{D^*_0}^{\prime \prime}}{x_{2} \hat{N}_{1}^{\prime} \hat{N}_{1}^{\prime \prime}}\left[x_{1}\left(M_{0}^{\prime 2}+M_{0}^{\prime \prime 2}\right)+x_{2} q^{2}\right.\non &&
		\left.-x_{2}\left(m_{1}^{\prime}+m_{1}^{\prime
			\prime}\right)^{2}-x_{1}\left(m_{1}^{\prime}-m_{2}\right)^{2}-x_{1}\left(m_{1}^{\prime
			\prime}+m_{2}\right)^{2}\right],\\
		F_{0}^{B D^*_0}\left(q^{2}\right) &=& F_{1}^{B D^*_0}\left(q^{2}\right)+\frac{q^{2}}{q \cdot P} \frac{N_{c}}{16 \pi^{3}} \int d x_{2} d^{2} p_{\perp}^{\prime}
		\frac{2 h_{B}^{\prime} h_{D^*_0}^{\prime \prime}}{x_{2} \hat{N}_{1}^{\prime} \hat{N}_{1}^{\prime \prime}}\left\{-x_{1} x_{2} M^{\prime 2}
		-p_{\perp}^{\prime 2}-m_{1}^{\prime} m_{2}\right.\non &&\left.
		-\left(m_{1}^{\prime \prime}+m_{2}\right)\left(x_{2} m_{1}^{\prime}+x_{1} m_{2}\right)
		+2 \frac{q \cdot P}{q^{2}}\left(p_{\perp}^{\prime 2}+2 \frac{\left(p_{\perp}^{\prime} \cdot q_{\perp}\right)^{2}}{q^{2}}\right)
		+2 \frac{\left(p_{\perp}^{\prime} \cdot q_{\perp}\right)^{2}}{q^{2}}\right.\non &&\left.
		-\frac{p_{\perp}^{\prime} \cdot q_{\perp}}{q^{2}}\left[M^{\prime \prime 2}-x_{2}\left(q^{2}+q \cdot P\right)
		-\left(x_{2}-x_{1}\right) M^{\prime 2}+2 x_{1} M_{0}^{\prime 2}\right.\right.\non &&\left.\left.-2\left(m_{1}^{\prime}-m_{2}\right)
		\left(m_{1}^{\prime}-m_{1}^{\prime \prime}\right)\right]\right\},\\
		A^{B\;^iD_{1}}(q^{2})&=&(M^{\prime}-M^{\prime\prime})\frac{N_{c}}{16 \pi^{3}} \int d x_{2} d^{2} p_{\perp}^{\prime} \frac{2 h_{B}^{\prime}
			h_{\;^iD_{1}}^{\prime \prime}}{x_{2} \hat{N}_{1}^{\prime} \hat{N}_{1}^{\prime \prime}}\left\{x_{2} m_{1}^{\prime}
		+x_{1} m_{2}+\left(m_{1}^{\prime}+m_{1}^{\prime \prime}\right) \frac{p_{\perp}^{\prime} \cdot q_{\perp}}{q^{2}}\right.\non &&\left.
		+\frac{2}{w_{^iD_{s1}}^{\prime \prime}}\left[p_{\perp}^{\prime 2}+\frac{\left(p_{\perp}^{\prime} \cdot q_{\perp}\right)^{2}}{q^{2}}\right]\right\},\\
		V_1^{B\;^iD_{1}}(q^{2})&=& -\frac{1}{M^{\prime}-M^{\prime\prime}}\frac{N_{c}}{16 \pi^{3}} \int d x_{2} d^{2} p_{\perp}^{\prime} \frac{h_{B}^{\prime} h_{\;^iD_{1}}^{\prime \prime}}{x_{2}
			\hat{N}_{1}^{\prime}
			\hat{N}_{1}^{\prime \prime}}\{2 x_{1}\left(m_{2}-m_{1}^{\prime}\right)\left(M_{0}^{\prime 2}+M_{0}^{\prime \prime 2}\right)
		+4 x_{1} m_{1}^{\prime \prime} M_{0}^{\prime 2}\non&&+2 x_{2} m_{1}^{\prime} q \cdot P
		\left.+2 m_{2} q^{2}-2 x_{1} m_{2}\left(M^{\prime 2}+M^{\prime \prime 2}\right)+2\left(m_{1}^{\prime}-m_{2}\right)\left(m_{1}^{\prime}
		-m_{1}^{\prime \prime}\right)^{2}+8\left(m_{1}^{\prime}-m_{2}\right) \right.\non &&
		\left. \times\left[p_{\perp}^{\prime 2}+\frac{\left(p_{\perp}^{\prime}
			\cdot q_{\perp}\right)^{2}}{q^{2}}\right]+2\left(m_{1}^{\prime}-m_{1}^{\prime \prime}\right)\left(q^{2}+q \cdot P\right) \frac{p_{\perp}^{\prime} \cdot q_{\perp}}{q^{2}}
		-4 \frac{q^{2} p_{\perp}^{\prime 2}+\left(p_{\perp}^{\prime} \cdot q_{\perp}\right)^{2}}{q^{2} w_{^iD_{s1}}^{\prime \prime}}
		\right.\non && \left.\times\left[2 x_{1}\left(M^{\prime 2}+M_{0}^{\prime 2}\right)-q^{2}-q \cdot P-2\left(q^{2}+q \cdot P\right) \frac{p_{\perp}^{\prime} \cdot q_{\perp}}{q^{2}}-2\left(m_{1}^{\prime}+m_{1}^{\prime \prime}\right)\left(m_{1}^{\prime}-m_{2}\right)\right]\right\},\;\;\;\;\;\;\;\\
		V_2^{B \;^iD_{1}}(q^{2})&=& (M^{\prime}-M^{\prime\prime})\frac{N_{c}}{16 \pi^{3}} \int d x_{2} d^{2} p_{\perp}^{\prime} \frac{2 h_{B}^{\prime} h_{\;^iD_{1}}^{\prime \prime}}{x_{2} \hat{N}_{1}^{\prime}
			\hat{N}_{1}^{\prime \prime}}\left\{(x_{1}-x_{2}\right)\left(x_{2} m_{1}^{\prime}+x_{1} m_{2}\right)-[2 x_{1} m_{2}
		-m_{1}^{\prime \prime}\non &&+\left(x_{2}-x_{1}\right) m_{1}^{\prime}]
		\times \frac{p_{\perp}^{\prime} \cdot q_{\perp}}{q^{2}}-2 \frac{x_{2} q^{2}+p_{\perp}^{\prime} \cdot q_{\perp}}{x_{2} q^{2} w_{^iD_{1}}^{\prime \prime}}[p_{\perp}^{\prime} \cdot p_{\perp}^{\prime \prime}
		+\left(x_{1} m_{2}+x_{2} m_{1}^{\prime}\right)\non &&\times\left(x_{1} m_{2}+x_{2} m_{1}^{\prime \prime}\right)]\},
	\end{eqnarray}
\end{footnotesize}
\begin{footnotesize}
	\begin{eqnarray}
		V_0^{B \;^iD_{1}}(q^{2})&=& \frac{M^{\prime}-M^{\prime\prime}}{2M^{\prime\prime}}V_1^{B \;^iD_{1}}(q^{2})-\frac{M^{\prime}+M^{\prime\prime}}{2M^{\prime\prime}}V_2^{B \;^iD_{1}}(q^{2})-\frac{q^2}{2M^{\prime\prime}}\frac{N_{c}}{16 \pi^{3}} \int d x_{2} d^{2} p_{\perp}^{\prime} \frac{h_{B}^{\prime} h_{\;^iD_{1}}^{\prime \prime}}{x_{2} \hat{N}_{1}^{\prime}
			\hat{N}_{1}^{\prime \prime}}\non &&\times\{2\left(2 x_{1}-3\right)\left(x_{2} m_{1}^{\prime}+x_{1} m_{2}\right)-8\left(m_{1}^{\prime}-m_{2}\right)
		\left[\frac{p_{\perp}^{\prime 2}}{q^{2}}
		+2 \frac{\left(p_{\perp}^{\prime} \cdot q_{\perp}\right)^{2}}{q^{4}}\right]-[\left(14-12 x_{1}\right) m_{1}^{\prime}\non &&+2 m_{1}^{\prime \prime}-\left(8-12 x_{1}\right) m_{2}] \frac{p_{\perp}^{\prime} \cdot q_{\perp}}{q^{2}}
		+\frac{4}{w_{^iD_{1}}^{\prime \prime}}(\left[M^{\prime 2}+M^{\prime \prime 2}-q^{2}+2\left(m_{1}^{\prime}-m_{2}\right)\left(-m_{1}^{\prime \prime}
		+m_{2}\right)\right]\non &&\times\left(A_{3}^{(2)}+A_{4}^{(2)}-A_{2}^{(1)}\right)
		+Z_{2}\left(3 A_{2}^{(1)}-2 A_{4}^{(2)}-1\right)+\frac{1}{2}[x_{1}\left(q^{2}+q \cdot P\right)
		-2 M^{\prime 2}-2 p_{\perp}^{\prime} \cdot q_{\perp}\non &&-2 m_{1}^{\prime}\left(-m_{1}^{\prime \prime}+m_{2}\right)
		\left.-2 m_{2}\left(m_{1}^{\prime}-m_{2}\right)\right]\left(A_{1}^{(1)}+A_{2}^{(1)}-1\right)\non &&
		\left.\left.\times q \cdot P\left[\frac{p_{\perp}^{\prime 2}}{q^{2}}
		+\frac{\left(p_{\perp}^{\prime} \cdot q_{\perp}\right)^{2}}{q^{4}}\right]\left(4 A_{2}^{(1)}-3\right)\right)\right\},\;\;\;
	\end{eqnarray}
\end{footnotesize}
with $i=1,3$.


\end{document}